\documentclass[twocolumn,resetfootnote,twocolappendix,floatfix]{aastex701}
\usepackage{amsmath}
\usepackage{lmodern}
\usepackage{placeins}
\usepackage{natbib}

\newcommand{\jkt}[2]{{#2}}
\usepackage[T1]{fontenc}
\usepackage{hyperref}
\usepackage{nameref}
\DeclareTextCompositeCommand{\.}{T1}{i}{\accent"0A \i}

\begin{document}

\title{{Chasing the Tides: Searching for Orbital Decay Signatures in Transit Timing Data and Tidal Models for 20 Hot Jupiters}}

\author[0009-0007-0502-7359]{Ahmet Cem Kutluay}
\affiliation{Astrophysics Group, Keele University, Staffordshire, ST5 5BG, UK}
\affiliation{Ankara University, Graduate School of Natural \& Applied Sciences, Astronomy \& Space Sciences Department, Ziraat Mahallesi {\.I}rfan Ba\c{s}tu\u{g} Caddesi, D{\i}\c{s}kap{\i}, TR-06110 Alt{\i}nda\u{g} / Ankara, T\"urkiye}
\affiliation{Ankara University, Astronomy and Space Sciences Research and Application Center (Kreiken Observatory), {\.I}ncek Blvd., TR-06837, Ahlatl{\i}bel, Ankara, T\"urkiye}
\email{a.c.kutluay@keele.ac.uk}

\author[0000-0002-4746-0181]{Özgür Baştürk}
\affiliation{Ankara University, Astronomy and Space Sciences Research and Application Center (Kreiken Observatory), {\.I}ncek Blvd., TR-06837, Ahlatl{\i}bel, Ankara, T\"urkiye}
\affiliation{Ankara University, Faculty of Science, Astronomy \& Space Sciences Department, Tando\u{g}an, TR-06100, Ankara, T\"urkiye}
\email{obasturk@ankara.edu.tr}

\author[0000-0003-4397-7332]{Adrian J. Barker}
\affiliation{School of Mathematics, University of Leeds, Leeds LS2 9JT, UK}
\email{A.J.Barker@leeds.ac.uk}

\author[0000-0002-5224-247X]{Sel\c{c}uk Yal\c{c}{\i}nkaya}
\affiliation{Ankara University, Graduate School of Natural \& Applied Sciences, Astronomy \& Space Sciences Department, Ziraat Mahallesi {\.I}rfan Ba\c{s}tu\u{g} Caddesi, D{\i}\c{s}kap{\i}, TR-06110 Alt{\i}nda\u{g} / Ankara, T\"urkiye}
\affiliation{Ankara University, Astronomy and Space Sciences Research and Application Center (Kreiken Observatory), {\.I}ncek Blvd., TR-06837, Ahlatl{\i}bel, Ankara, T\"urkiye}
\affiliation{Ankara University, Faculty of Science, Astronomy \& Space Sciences Department, Tando\u{g}an, TR-06100, Ankara, T\"urkiye}
\affiliation{Astrobiology Research Unit, Universit\'e de Li\`ege, All\'ee du 6 Ao\^ut 19C, B-4000 Li\`ege, Belgium}
\email{selcuk_yalcinkaya@yahoo.com}

\author[0000-0002-3807-3198]{John Southworth}
\affiliation{Astrophysics Group, Keele University, Staffordshire, ST5 5BG, UK}
\email{taylorsouthworth@gmail.com}

\author[0000-0002-4953-4818]{Sel{\.{i}}m Osman Selam}
\affiliation{Ankara University, Astronomy and Space Sciences Research and Application Center (Kreiken Observatory), {\.I}ncek Blvd., TR-06837, Ahlatl{\i}bel, Ankara, T\"urkiye}
\affiliation{Ankara University, Faculty of Science, Astronomy \& Space Sciences Department, Tando\u{g}an, TR-06100, Ankara, T\"urkiye}
\email{selam@science.ankara.edu.tr}

\author[0009-0007-8172-6602]{\"Ozlem \c{S}{\.i}m\c{s}{\.i}r}
\affiliation{Ankara University, Graduate School of Natural \& Applied Sciences, Astronomy \& Space Sciences Department, Ziraat Mahallesi {\.I}rfan Ba\c{s}tu\u{g} Caddesi, D{\i}\c{s}kap{\i}, TR-06110 Alt{\i}nda\u{g} / Ankara, T\"urkiye}
\affiliation{Ankara University, Astronomy and Space Sciences Research and Application Center (Kreiken Observatory), {\.I}ncek Blvd., TR-06837, Ahlatl{\i}bel, Ankara, T\"urkiye}
\email{ozsmsr@gmail.com}

\author[0000-0003-0631-1961]{Kaan Kaplan}
\affiliation{Sivas Science and Technology University, G\"ultepe Neighborhood, Mecnun Otyakmaz Street No:1, Sivas}
\email{kplnkaan@gmail.com}

\author[0000-0003-4419-2908]{Furkan Akar}
\affiliation{Ankara University, Graduate School of Natural \& Applied Sciences, Astronomy \& Space Sciences Department, Ziraat Mahallesi {\.I}rfan Ba\c{s}tu\u{g} Caddesi, D{\i}\c{s}kap{\i}, TR-06110 Alt{\i}nda\u{g} / Ankara, T\"urkiye}
\affiliation{Ankara University, Astronomy and Space Sciences Research and Application Center (Kreiken Observatory), {\.I}ncek Blvd., TR-06837, Ahlatl{\i}bel, Ankara, T\"urkiye}
\email{furkanakar_@outlook.com}

\author[0009-0000-8026-2104]{İpek Aleyna Ert\"urk}
\affiliation{Ankara University, Graduate School of Natural \& Applied Sciences, Astronomy \& Space Sciences Department, Ziraat Mahallesi {\.I}rfan Ba\c{s}tu\u{g} Caddesi, D{\i}\c{s}kap{\i}, TR-06110 Alt{\i}nda\u{g} / Ankara, T\"urkiye}
\affiliation{Ankara University, Astronomy and Space Sciences Research and Application Center (Kreiken Observatory), {\.I}ncek Blvd., TR-06837, Ahlatl{\i}bel, Ankara, T\"urkiye}
\email{ipekaerturk@hotmail.com}

\author[0009-0002-8061-8857]{Zeynep Zeng{\.i}n}
\affiliation{Middle East Technical University, Physics Department, 06800, Ankara, T\"urkiye}
\email{zeynep.zengin_01@metu.edu.tr}

\author[0009-0009-5717-5382]{Ebrar Akal{\i}n}
\affiliation{Middle East Technical University, Physics Department, 06800, Ankara, T\"urkiye}
\email{ebrarakalin@gmail.com}

\author[0009-0001-8728-9200]{Volkan \"Ozsoy}
\affiliation{Middle East Technical University, Physics Department, 06800, Ankara, T\"urkiye}
\email{volkanozsoy0707@gmail.com}

\author[0009-0007-6688-5584]{\"Ozdenur Yald{\i}r}
\affiliation{Middle East Technical University, Physics Department, 06800, Ankara, T\"urkiye}
\email{ozdenuryaldir@gmail.com}

\author[0009-0008-2515-5830]{D{\.i}lara {\.I}\c{c}\"oz}
\affiliation{Middle East Technical University, Physics Department, 06800, Ankara, T\"urkiye}
\email{dilara.icoz@metu.edu.tr}

\author[0000-0002-9428-8732]{Luigi Mancini}
\affiliation{Department of Physics, University of Rome ``Tor Vergata'', Via
della Ricerca Scientifica 1, 00133 Rome, Italy}
\affiliation{INAF -- Turin Astrophysical Observatory, via Osservatorio 20, 10025 Pino Torinese, Italy}
\email{lmancini@roma2.infn.it}

\author[0009-0003-2191-8873]{Burak Duru}
\affiliation{Ankara University, Graduate School of Natural \& Applied Sciences, Astronomy \& Space Sciences Department, Ziraat Mahallesi {\.I}rfan Ba\c{s}tu\u{g} Caddesi, D{\i}\c{s}kap{\i}, TR-06110 Alt{\i}nda\u{g} / Ankara, T\"urkiye}
\affiliation{Ankara University, Astronomy and Space Sciences Research and Application Center (Kreiken Observatory), {\.I}ncek Blvd., TR-06837, Ahlatl{\i}bel, Ankara, T\"urkiye}
\email{duruburak.tr@gmail.com}

\author[0000-0001-5715-1166]{Fatma Tezcan}
\affiliation{Türkiye National Observatories, DAG, 25050, Erzurum, Türkiye}
\affiliation{Atatürk University, Astronomy and Astrophysics, Erzurum, Türkiye}
\email{fatmat@trgozlemevleri.gov.tr}

\author[0009-0007-9111-5629]{Alk{\i}m {\"O}zf{\.i}dan}
\affiliation{Ankara University, Graduate School of Natural \& Applied Sciences, Astronomy \& Space Sciences Department, Ziraat Mahallesi {\.I}rfan Ba\c{s}tu\u{g} Caddesi, D{\i}\c{s}kap{\i}, TR-06110 Alt{\i}nda\u{g} / Ankara, T\"urkiye}
\affiliation{Ankara University, Astronomy and Space Sciences Research and Application Center (Kreiken Observatory), {\.I}ncek Blvd., TR-06837, Ahlatl{\i}bel, Ankara, T\"urkiye}
\email{alkozfidan@gmail.com}

\author[0009-0007-6246-0933]{Ushna Umar}
\affiliation{Department of Physics, Faculty of Science, Bilkent University, TR-06800 Ankara, T\"urkiye}
\email{ushna.umar@ug.bilkent.edu.tr}

\author[0000-0002-6176-9847]{Ana\"{e}l W\"unsche}
\affiliation{Baronnies Provençales Observatory, Hautes Alpes, Parc Naturel Regional des Baronnies Provençales, F-05150 Moydans, France}
\email{anael.wunsche@obs-bp.fr}

\author[0000-0002-5854-4217]{Martin J. Burgdorf}
\affiliation{Earth System Sciences, Atmospheric Science, University of Hamburg, Hamburg, Germany}
\email{martin.burgdorf@uni-hamburg.de}

\author[0009-0007-5946-8731]{Richard E. Cannon}
\affiliation{Institute for Astronomy, University of Edinburgh, Royal Observatory, Edinburgh EH9 3HJ, UK}
\email{richard.cannon@ed.ac.uk}

\author[0000-0003-3425-6605]{Roberto Jose Figuera Jaimes}
\affiliation{Instituto de Astronom\'{i}a y Ciencias Planetarias, Universidad de Atacama, Copiap\'{o} 485, Copiapó, Chile}
\affiliation{Millennium Institute of Astrophysics MAS, Nuncio Monsenor Sotero Sanz 100, Of. 104, Providencia, Santiago, Chile}
\affiliation{Instituto de Astrofísica, Facultad de Física, Pontificia Universidad Católica de Chile, Av. Vicuña Mackenna 4860, 7820436, Macul, Santiago, Chile}
\affiliation{Centre for Exoplanet Science, SUPA, School of Physics \& Astronomy, University of St Andrews, North Haugh, St Andrews KY16 9SS, UK}
\email{robertofiguera@gmail.com}

\author[0000-0001-8870-3146]{Tobias Cornelius Hinse}
\affiliation{University of Southern Denmark, Department of Physics, Chemistry and Pharmacy, SDU-Galaxy, Campusvej 55, 5230, Odense M, Denmark}
\email{toch@cp3.sdu.dk}

\author[]{Vincent Okoth}
\affiliation{Institute for Astronomy, University of Edinburgh, Royal Observatory, Edinburgh EH9 3HJ, UK}
\email{}

\author[0000-0002-9024-4185]{Jeremy Tregloan-Reed}
\affiliation{Centro de Astronomía, Universidad de Antofagasta, Av. Angamos 601, Antofagasta, Chile}
\email{jeremy.tregloanreed@uantof.cl}

\author[0009-0007-5714-2391]{El{\.i}f S{\i}la Bu\u{g}day}
\affiliation{Department of Physics, Faculty of Science, Bilkent University, TR-06800 Ankara, T\"urkiye}
\email{sila.bugday@ug.bilkent.edu.tr}

\author[0009-0001-3015-7426]{Utku Akdere}
\affiliation{Department of Physics, Faculty of Science, Bilkent University, TR-06800 Ankara, T\"urkiye}
\email{utku.akdere@ug.bilkent.edu.tr}

\author[0009-0003-9028-3584]{Y{\.i}{\u{g}}{\.i}t Turan}
\affiliation{Department of Physics, Imperial College London, London, SW7 2AZ, UK}
\email{yigit.turan25@imperial.ac.uk}

\author[0000-0002-6990-8899]{S{\.i}nan Al{\.i}\c{s}}
\affiliation{Department of Astronomy and Space Sciences, Faculty of Science, Istanbul University, 34119 {\.{I}}stanbul, T\"urkiye}
\email{salis@istanbul.edu.tr}

\author[0000-0003-2253-9499]{C{\.i}han Tu\c{g}rul Tezcan}
\affiliation{Türkiye National Observatories, DAG, 25050, Erzurum, Türkiye}
\email{cihant@trgozlemevleri.gov.tr}

\author[0000-0003-2675-3564]{Fuat Korhan Yelkenc{\.i}}
\affiliation{Department of Astronomy and Space Sciences, Faculty of Science, Istanbul University, 34119 {\.{I}}stanbul, T\"urkiye}
\email{yelkenci@istanbul.edu.tr}

\author[]{Selina Hajarat}
\affiliation{Department of Physics, Faculty of Science, Bilkent University, TR-06800 Ankara, T\"urkiye}
\email{selina.hajarat@ug.bilkent.edu.tr}

\begin{abstract}
In this work, we present a transit timing variation analysis for 20 hot Jupiter systems, which we interpret with 
theoretical tidal dissipation models. For the majority of the sample, we conclude that a constant orbital period model represents the timing data best. Only WASP-12\,b, TrES-1\,b and WASP-121\,b exhibit a changing orbital period, according to the most up-to-date results. We updated the orbital decay rate of WASP-12\,b to $\dot{P} = -29.4 \pm 4.0 {\rm ~ms~yr^{-1}}$ and the corresponding stellar tidal quality factor to $Q_*^{\prime} = 1.72 \pm 0.18 \times 10^5$. For TrES-1\,b, the median quadratic model suggests a period decrease at a rate of $-14.9 \pm 0.6 {\rm ~ms~yr^{-1}}$, but the corresponding $Q_*^{\prime} = 570 \pm 60$ does not agree with the theoretical estimates, which suggest $Q_*^{\prime} \sim 10^6$ due to internal gravity wave dissipation. Lastly, WASP-121\,b exhibits orbital growth at a rate of $15.1 \pm  0.8 {\rm ~ms~yr^{-1}}$, and theoretical results support outward migration due to strong inertial wave dissipation.
\end{abstract}

\keywords{
\uat{Exoplanets}{498},
\uat{Hot Jupiters}{753},
\uat{Transit timing variation method}{1710},
\uat{Tidal interaction}{1699},
\uat{Star-planet interactions}{2177}
}


\section{Introduction} \label{sec:introduction}

Tidal interactions are key driving mechanism of orbital and rotational evolution in both binary stars and exoplanetary systems, especially for planets with short orbital periods. These interactions lead to a transfer of angular momentum between the planet and its host star via energy dissipation, leading to orbital circularization, tidal synchronization, or, in some cases, orbital decay \citep[e.g.][]{hut1980,hut1981,ogilvie2014,B2025}. Circularization of hot Jupiters (hereafter HJs) is primarily driven by dissipation within the planet \citep[e.g.][]{guillot1996,penev2024,Lazovik2024}, while stellar tides mostly dominate orbital decay for planets orbiting faster than the stellar rotation. Once circularized, further tidal dissipation in the star continuously removes angular momentum from the orbit, causing close-in planets to migrate inward over long timescales 
\citep[e.g.][]{essickweinberg2016,Chernov2017,barker2020}, such as the several Myr decay timescale inferred for WASP-12\,b \citep[e.g.][]{maciejewski2016,patra2017,kutluay2023}.

HJs provide a natural laboratory for studying tidal theory in star-planet systems. Their short orbital periods ($P_\mathrm{orb} < 10$~d; \citeauthor{wang2015}~\citeyear{wang2015}) and proximity to their host stars enhance the magnitude of tidal forces. 
Two pieces of evidence indicating the efficiency of tides in these systems include: (1) the predominance of circular orbits among the shortest-period systems \citep[e.g.][]{penev2024}; and (2) the relatively low velocity scatter of HJ host stars, which is an indicator that they represent a younger population potentially evolving further through tidally-driven inspiral \citep[e.g.][]{hamerschlaufman2019}. In addition to that, direct observational evidence for orbital decay of HJs has been provided by recent JWST observations of ZTF SLRN-2020 \citep{de2023,lau2025}.

A comprehensive picture of orbital decay comes from two complementary approaches. The orbital period of a planet changes as its orbit shrinks (or grows) as a result of angular momentum exchanges. 
Transit Timing Variations (TTVs) provide a powerful method for detecting variations in transit mid-times on the order of seconds to minutes over long timescales \citep{agol2005}. Identifying such small deviations requires a long temporal baseline of transit observations \citep{patra2017,winn2019}, as longer high-precision datasets increase sensitivity to gradual changes in the orbital period and improve model discrimination \citep[see Eq. 35 in][]{jackson2023}. Even tiny TTV shifts accumulate over years, potentially providing measurable signatures of orbital decay. Once statistically significant changes in orbital period are detected, the efficiency of tidal dissipation in the host star can be estimated by constraining the modified tidal quality factor ($Q^\prime$), defined as the ratio of the tidal potential energy stored ($E_0$) to the kinetic energy dissipated per oscillation period $(\oint -\dot{E} \, dt$) due to tidal interactions \citep[e.g.][]{goldreichsoter1966,ogilvie2014}:
\begin{equation}
Q' = \frac{3Q}{2k_2} = \frac{3}{2k_2} 2\pi E_0 \left( \oint -\dot{E} \, dt \right)^{-1}.
\end{equation}
Here $k_2$ is the quadrupolar Love number, which measures the degree of central condensation of a body, and takes a small value $\lesssim 0.05$ for a solar-type star. We have incorporated it into the definition of the tidal quality factor, as it is $Q'$ that appears in tidal evolutionary equations.

If the rate of orbital period change due to tidal interactions can be determined, the stellar tidal quality factor ($Q^\prime_{\star}$) can also be calculated using observable parameters according to \citep[e.g.][]{Birkby2014,Wilkins2017,patra2017,mancini2022}
\begin{equation}
Q^\prime_{\star} = -\frac{27}{2} \pi \left( \frac{M_p}{M_\star} \right) \left( \frac{a}{R_\star} \right)^{-5} \left( \frac{dP_{\text{orb}}}{dE} \right)^{-1} P_{\text{orb}}.
\label{eq:modified_tidal_quality_factor}
\end{equation}
Here $P_{\mathrm{orb}}$ is the orbital period, $a$ is the semi-major axis, $M_{\rm p}$ and $M_\star$ are the planet and star masses respectively, $R_\star$ is the stellar radius, and $dP_{\mathrm{orb}}/dE$ is the orbital period change per cycle. 

This method constrains the tidal quality factor at a snapshot in time, but $Q_\star^\prime$ is generally not constant and depends on stellar properties that \jkt{evolve}{change as the star evolves}. Evolved main-sequence stars and subgiants tend to exhibit more efficient tidal dissipation, corresponding to lower $Q_\star^\prime$ values \citep[e.g.][]{mathis2015,gallet2017,barker2020,Esseldeurs2024}.
Secondly, tidal dissipation is not believed to be governed by a single, uniform mechanism. Instead, depending on the region of a star and stellar evolutionary phase, different mechanisms can contribute to dissipation with varying levels of efficiency \citep[e.g.][]{barker2020}. Recognizing the variability in tidal dissipation mechanisms across different stellar regions and evolutionary stages prompts a deeper examination of the underlying physical processes, which include equilibrium tides, inertial waves, and internal gravity waves.

\textsl{Equilibrium tides} describe the large-scale quasi-hydrostatic deformation of a body. For a quadrupolar tide ($l=2$, $m=2$), this is a spheroidal bulge and its associated flow \citep{ogilvie2014}. In stellar convective zones, turbulent convection dissipates the flow's kinetic energy \citep[e.g.][]{ogilvielin2007,duguid2020}. For HJ systems, equilibrium tides are unlikely to cause significant orbital decay during the main sequence, with $Q_{\rm eq}^\prime \gtrsim 10^8$ for $0.8$--$1.4\,M_\odot$ stars, giving orbital decay timescales $\gtrsim 100$~Gyr, far exceeding stellar lifetimes \citep{penevsasselov2011,barker2020}.

\textsl{Inertial waves} (IW) are excited when the tidal forcing frequency lies in $[-2\Omega, 2\Omega]$, where $\Omega$ is the stellar rotation rate \citep[e.g.][]{ogilvielin2007,ogilvie2014,astoul2023}. Restored by Coriolis forces, they propagate through the convective envelope and are most relevant in low-mass stars with deep convection zones. In HJ systems, IW rarely produce significant orbital decay during the main sequence, as the frequency condition is usually not satisfied for their excitation, though it is more likely to be satisfied in young or tidally spun-up stars. The resulting tidal quality factors span $Q^\prime_\star \sim 10^7$--$10^{12}(P_\mathrm{rot}/10\mathrm{d})^2$ for $1.0$--$1.6\,M_\odot$ stars (where $P_\mathrm{rot}=2\pi/\Omega$ is the stellar rotation period), indicating modestly efficient to inefficient dissipation. As stars evolve into subgiants, the expanding convective envelope and contracting radiative core enhance tidal dissipation, enabling IW excitation with $Q^{'}_{\star} \sim 10^4$--$10^5 (P_\mathrm{rot}/10\mathrm{d})^2$ \citep{barker2020}. 


\textsl{Internal gravity waves} (IGW) are generated by tidal forces at the convective--radiative interface \citep[e.g.][]{zahn1975,zahn1977} and are restored by buoyancy forces. They propagate toward the stellar core, where they can amplify and break, leading to efficient dissipation \citep[e.g.][]{goodmandickson1998,ogilvielin2007,barkerogilvie2010,barker2011,weinberg2012}. The efficiency of this process depends on the wave amplitude (see Eq.~47 in \citeauthor{barker2020} \citeyear{barker2020}). In particular, once a critical amplitude is exceeded, wave breaking produces efficient dissipation with a tidal quality factor $Q^\prime_\star \sim 10^5 (P_\mathrm{tide}/0.5 \,\mathrm{d})^{8/3}$ (where $P_\mathrm{tide}$ is the tidal period) in main-sequence stars \citep{barkerogilvie2010,barker2011}. This critical amplitude can be interpreted in terms of a critical planetary mass (Eq.~50 in \citealt{barker2020}), linking IGW breaking directly to planetary properties.

Several studies have highlighted the importance of IGWs for the orbital evolution of planetary systems \citep[e.g.][]{Chernov2017,weinberg2017,barker2020,Ahuir2021,MF2021,Guo2023}, especially for host stars near the end of the main sequence, when short-period HJs are expected to be engulfed as their stars evolve \citep{hamerschlaufman2019,hamerschlaufman2020,mustill2021,weinberg2024}.\par
An alternative mechanism for energy dissipation and outward propagation has been proposed for F-type stars with masses $1.2$--$1.6\,M_{\odot}$. In this scenario, IGWs propagating toward the core can be converted into magnetic waves by the strong magnetic field generated by the convective-core dynamo, provided the field is sufficiently strong \citep{duguid2024}. 


Formulas to simply evaluate the dissipation caused by each of these mechanisms \jkt{have been written in the literature}{are available} \citep[e.g.][]{barker2020}. The modified tidal quality factor for inertial waves can be obtained by
\begin{equation}
\frac{1}{\left\langle Q_{\text{IW}}^{\prime} \right\rangle} = \frac{32 \pi^2 G}{3 (2l + 1) R^{2l + 1} |A|^2} \left( E_l + E_{l-1} + E_{l+1} \right),
\label{eq:IW}
\end{equation}
where $E_l$, $E_{l-1}$ and $E_{l+1}$ are parameters directly calculated from stellar evolution models, as described in Eqs. 31, 32 and 33 of \cite{barker2020}. It should be noted that this formalism follows the frequency-averaged dissipation of inertial waves proposed by \citep{ogilvie2013}. 
The modified tidal quality factor for IGWs is:
\begin{equation}
\frac{1}{Q_{\text{IGW}}^{\prime}} = \frac{2 \left[ \Gamma\left( \frac{1}{3} \right) \right]^2}{3^{1/3} (2l+1) \left( l(l+1) \right)^{4/3}}\frac{R}{G {M_\star^2}} \mathcal{G} |\omega|^{8/3},
\label{eq:IGW}
\end{equation}

\begin{figure*}
    \centering
    \includegraphics[width=\textwidth]{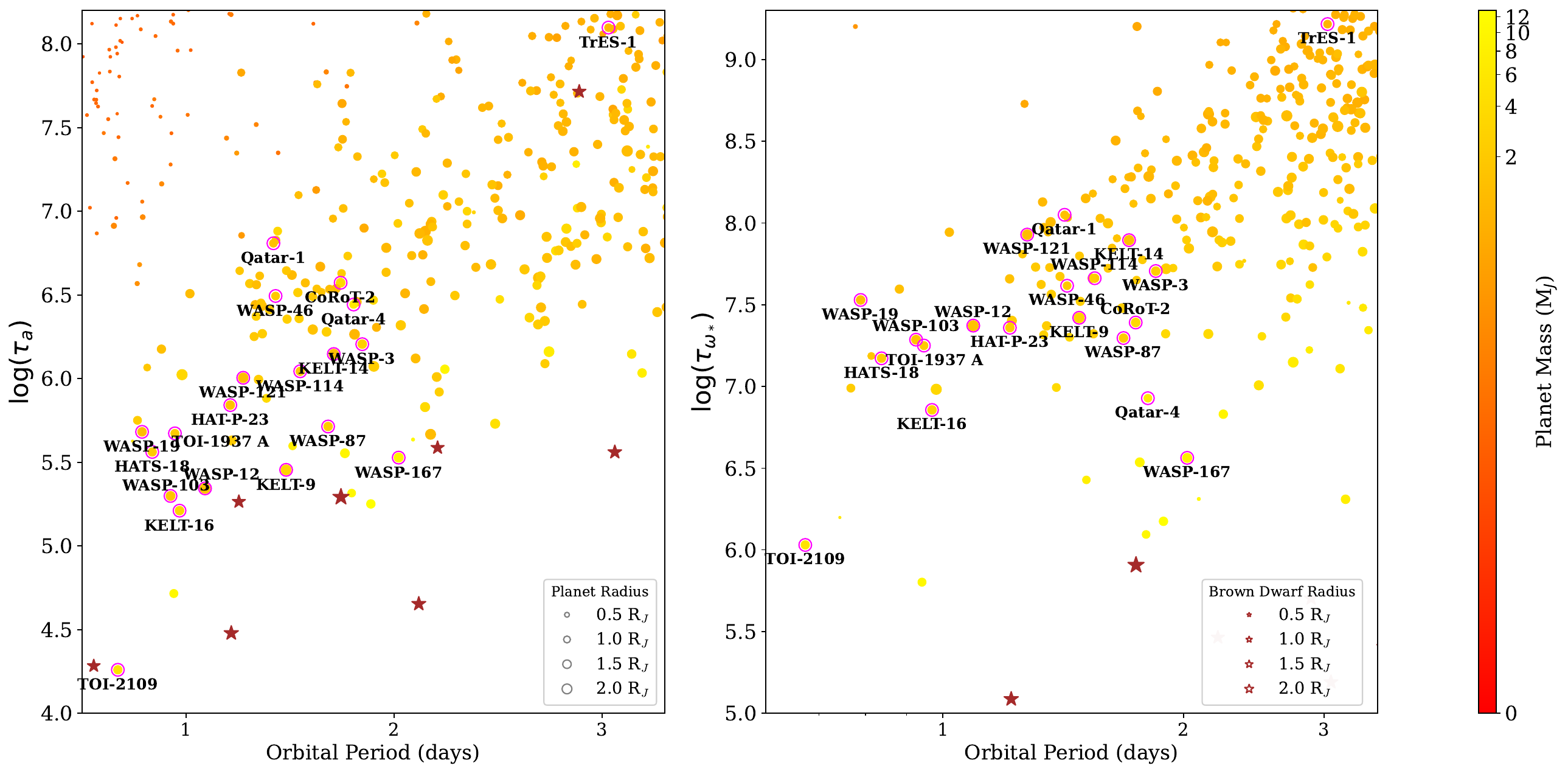}
    \caption{Orbital period versus the selection criteria parameters $\tau_a$ \jkt{\&}{and} $\tau_{\omega_\star}$ for \jkt{the catalogue}{objects} from NASA Exoplanet Archive and TEPCat. The symbol size is proportional to the planetary (or brown dwarf) radius, while the color scale represents the planetary mass. The systems analyzed in this study are highlighted and labeled.}
    \label{fig:tau_comparison}
\end{figure*}

\noindent where the quantity \(\mathcal{G}\) represents a combination of stellar structure parameters that characterize the properties of the radiative-convective boundary in a given stellar model (Eq.~42 from \citeauthor{barker2020} \citeyear{barker2020}) and $\omega = 2(\Omega_o - \Omega_s) =2\pi/P_\mathrm{tide}$ is the tidal frequency for a circular, aligned orbit. We do not provide a formulation for equilibrium tides as they are less important than IWs and IGWs for HJs based on our current theoretical understanding.

If all or a combination of these processes operate, their effects can be summed via \citep{lazovik2021}
\begin{equation}
\frac{1}{Q^{\prime}} = \frac{1}{Q_{\text{eq}}^{\prime}} + \frac{1}{Q_{\text{IW}}^{\prime}} + \frac{1}{Q_{\text{IGW}}^{\prime}}.
\end{equation}
The resulting dissipation from tidal forcing can leave a detectable signature in the orbital period variations observed through TTV analysis, providing insights into the stellar evolutionary phase, internal structure, and the timescales of orbital evolution.

Numerous studies have suggested that orbital decay signals can be detected observationally. As an example, the orbital decay of WASP-18\,b was predicted to have become observable within a decade, with a reported decay timescale of $1 \times 10^6$\jkt{ years}{~yr} assuming $Q_\star^{\prime} = 10^6$ \citep{hellier2009}, though this has not been detected to date \citep[e.g.][]{Wilkins2017}. Similarly, \citet{penev2016} proposed that HATS-18\,b should have undergone orbital decay due to stellar tidal dissipation, a claim later reinforced by 
\citet{penev2018}, who found $\log Q_\star^\prime = 7.18_{-0.173}^{+0.205}$ (but see also \citealt{southworth2022}). Additional cases include KELT-16\,b \citep{maciejewski2018, mancini2022, harre2023, alvarado2024}, WASP-19\,b \citep{korth2023, biswas2024}, WASP-4\,b \citep{bouma2019,Me+09mn2,turner2022,basturk2025}, and WASP-43\,b \citep{hellier2011, hoyer2016}, all suggesting observable orbital decay signatures.

With this study, our aim is to investigate the potential orbital decay of 20 HJs through TTV analysis, supported by theoretical predictions of tidal dissipation from stellar models. The systems we study are: CoRoT-2, HAT-P-23, HATS-18, KELT-9, KELT-16, Qatar-1, Qatar-4, TOI-2109, TOI-1937 A, TrES-1, WASP-3, WASP-12, WASP-19, WASP-46, WASP-87, WASP-103, WASP-114, WASP-121, WASP-122  (KELT-14) and WASP-167.

The remainder of this paper is organised as follows. In \S~\ref{sec:target_selection}, we describe our target selection strategy. \S~\ref{sec:Data} presents the observational data and outlines the data reduction procedures. \S~\ref{sec:analysis_results} discusses the methodology we employ in our analysis and the system-by-system results. \S~\ref{sec:discussion} discusses our results further and finally, \S~\ref{sec:conclusion} summarizes our main conclusions.

\section{Target Selection}
\label{sec:target_selection}

The efficiency of tidal interactions depends strongly on fundamental system parameters, such as the stellar and planetary masses ($M_{\star}$ and $M_{\rm p}$), stellar radius ($R_\star$) and the orbital semi-major axis ($a$). These dependencies are partly encapsulated in the modified tidal quality factor (Eq.~\ref{eq:modified_tidal_quality_factor}). Moreover, as shown in Eqs. \ref{eq:IW} and \ref{eq:IGW}, the efficiency of tidal dissipation exhibits a steep dependence on the stellar radius, scaling as $R_\star^5$ in the quadrupolar ($l=2$) case, which underscores the dominant role of stellar structure and stellar evolutionary phase in tidal energy dissipation. Even for $Q^{'}_\star=$ const, independent of radius, tidal timescales tend to depend on stellar radius to a strong power, as we will show later with Eqs. \ref{eq:tau_a} and \ref{eq:tau_s}. \par
When selecting systems that are most likely to exhibit observable orbital decay due to tidal effects, we prioritize the physical parameters mentioned above, which naturally lead us to focus on close-in giant planets orbiting solar-like stars (with $P_{\rm orb} \lesssim 2$~d) — particularly those approaching the end of their main-sequence lifetimes, as IWs and IGWs tend to become more efficient as their host stars evolve (for fixed $P_\mathrm{rot}$ and $P_\mathrm{tide}$).

In this context, we employed two dimensionless metrics to estimate the relative strength of tidal interactions and their long-term dynamical effects, and thereby to select our best candidate systems. These \jkt{dimensionless metrics}{} were originally proposed by \cite{siverd2012} and are defined as follows:
\begin{equation}
\tau_a \equiv \left( \frac{M_{\star}}{M_{\rm p}} \right) \left( \frac{a}{R_\star} \right)^5,
\label{eq:tau_a}
\end{equation}
and
\begin{equation}
\tau_{\omega_\star} \equiv \left( \frac{M_{\star}}{M_{\rm p}} \right)^2 \left( \frac{a}{R_\star} \right)^3.
\label{eq:tau_s}
\end{equation} 
Note that these are dimensionless quantities and are not timescales themselves. If we adopt a constant time-lag tidal model
\citep{hut1981,matsumura2010} and assume a circular ($e = 0$), aligned orbit 
(the angle between the axes of stellar rotation and planetary orbit, $\lambda=0$ ) around a slowly rotating star, the characteristic timescale of orbital decay to within an O(1) factor (that we omit here and depends on the power of $a$ in the expression for $\dot{a}$) is $t_{\rm decay} \equiv a/|\dot{a}| = (12\pi)^{-1} Q_\star^\prime \tau_a P_{\rm orb}$ and the timescale for the host star to be synchronized with its orbiting planet is $t_{\rm sync} \equiv \omega_\star/|\dot{\omega_\star}| \propto Q_\star^\prime \tau_{\omega_\star} P_{\rm orb}$ \citep{matsumura2010}. Hence $\tau_a$ and $\tau_{\omega_\star}$ are good measures of the relative strengths of tidal interactions, although they do not incorporate the variation in $Q^{'}_\star$.

\begin{deluxetable*}{lcccccccclcc}
\tablecaption{Fundamental stellar and planetary parameters for selected HJ systems.}
\label{tab:exoplanet_systems}
\tablewidth{0pt}
\tablehead{
\colhead{System} & \colhead{$T_\mathrm{eff}$ (K)} & \colhead{Age (Gyr)} & \colhead{$M_\star$ ($M_\odot$)} & \colhead{$R_\star$ ($R_\odot$)} & & \colhead{$P_\mathrm{orb}$ (d)} & \colhead{$M_{\rm p}$ ($M_\mathrm{J}$)} & \colhead{$a/R_\star$} & \colhead{$\log(\tau_a)$} & \colhead{$\log(\tau_{\omega_\star})$} & \colhead{Ref.}
}
\startdata
    CoRoT-2     & $5625 \pm 120$  & $2.7^{+3.2}_{-2.7}$ & $0.970\pm0.060 $ & $0.902\pm0.018$ &  & $1.74299673(31)$ & $3.30^{+0.19}_{-0.18}$ & $6.70\pm0.19$ & $6.573$ & $7.391$ & 1,2 \\
    HAT-P-23    & $5905\pm80$ & $4\pm1$ & $1.130\pm0.035 $ & $1.203\pm0.074$  & & $1.21288287(17)$  & $1.97\pm0.12 $ & $4.15\pm0.26$ & $5.842$ & $7.359$ & 1 \\
    HATS-18     & $5600 \pm 120$ & $4.2\pm2.2$ & $1.037\pm0.047$ & $1.020^{+0.057}_{-0.031}$ & & $0.83784340(47)$ & $1.980\pm0.077$  & $3.71^{+0.11}_{-0.22}$ & $5.562$ & $7.172$ & 3 \\
    KELT-9 & $10170\pm450$  & $0.3$  &  $2.52^{+0.25}_{-0.20}$  & $2.362^{+0.075}_{-0.063}$ & & $1.4811235(11)$ & $2.88\pm0.84$ & $3.153\pm0.011$ & $5.455$ & $7.420$ & 4 \\
    KELT-16 & $6236\pm54$ & $3.1\pm0.3$ & $1.211^{+0.043}_{-0.046}$ & $1.360^{+0.064}_{-0.053}$ & & $0.9689951(24)$ & $2.75^{+0.16}_{-0.15}$ & $3.23^{+0.12}_{-0.13}$ & $5.211$ & $6.856$ & 5 \\
    Qatar-1 & $4910\pm100$ & $11.60^{+0.60}_{-4.70}$ & $0.850\pm0.030$ & $0.800\pm0.050$ & & $1.42002420(22)$ & $1.294^{+0.052}_{-0.049}$ & $6.247^{+0.067}_{-0.065}$ & 6.809 & 8.05 & 1,7 \\
    Qatar-4 & $5215\pm50$ & $0.17\pm0.01 $ & $0.896\pm0.048$ & $0.849\pm0.063$ & & $1.8053663(15)$ & $6.10\pm0.54$ & $7.11\pm0.48 $ & 6.443  & 6.928 & 8,9 \\
    TOI-2109    & $6540\pm160$ & $1.77^{+0.88}_{-0.68}$ & $1.453\pm0.074 $ & $1.698^{+0.062}_{-0.057}$ & & $0.672474140(28)$ & $5.02\pm0.75$ & $2.268\pm0.021$ & 4.26  & 6.03 & 10 \\
    TOI-1937 A  & $5814^{+91}_{-93}$ & $3.6^{+3.1}_{-2.3}$ & $1.072^{+0.059}_{-0.064}$ & $1.080^{+0.025}_{-0.024}$ & & $0.94667944(47)$ & $2.01^{+0.17}_{-0.16}$ & $3.85^{+0.09}_{-0.10}$& 5.673 & 7.25 & 11 \\
    TrES-1  & $5230\pm50$ & $3.7^{+3.4}_{-2.8}$ & $0.878^{+0.038}_{-0.040}$ & $0.807 ^{+0.017}_{-0.016}$ & & $3.03006973(18)$ & $0.752^{+0.047}_{-0.046}$ & $10.52^{+0.02}_{-0.18}$ & $8.098$ & $9.217$ & 1,12 \\
    WASP-3  & $6400\pm100$ & $2.1\pm1.2$ & $1.260\pm0.100$ & $1.366\pm0.044$ & &
    $1.84683510(40)$ & $1.89\pm0.12$ & $5.18\pm0.35$ & 6.206 & 7.706 & 1,23 \\
    WASP-12  & $6360^{+130}_{-140}$ & $3.05 \pm 0.32$ & $1.434^{+0.110}_{-0.090}$ & $1.657^{+0.046}_{-0.044}$ & & $1.09142030(14)$ & $1.470^{+0.076}_{-0.069}$ & $3.039^{+0.034}_{-0.033}$ & $5.344$ & $7.372$ & 1,7 \\
    WASP-19 & $5616^{+66}_{-65}$ & $6.40^{+4.10}_{-3.50}$ & $0.965^{+0.091}_{-0.095}$ & $1.006^{+0.031}_{-0.034}$ & & $0.78883852(82)$ & $1.154^{+0.078}_{-0.080}$ & $3.533^{+0.048}_{-0.052}$ & $5.682$ & $7.529$ & 1,13 \\
    WASP-46 & $5600\pm150$ & $9.60^{+3.40}_{-4.20}$ & $0.828\pm0.067$ & $0.858\pm0.024$ & & $1.43036763(93)$ & $1.91 \pm 0.13$ & $5.851^{+0.038}_{-0.037}$ & 6.494 &  7.616 & 1,15,25 \\
    WASP-87 & $6450\pm120$ & -- &  $1.204\pm0.093$ & $1.627\pm0.062$ & & $1.6827950(19)$ & $2.18\pm0.15$ & $3.89^{+0.49}_{-0.42}$ & 5.714 & 7.296 & 16,17,18 \\
    WASP-103 & $6110\pm160$ & $4\pm1$ & $1.220^{+0.039}_{-0.036}$ & $1.220^{+0.039}_{-0.036}$ & & $0.925542(19)$ & $1.490\pm0.088$ & $2.978^{+0.050}_{-0.096}$ & 5.299 & 7.286 & 19 \\
    WASP-114 & $5940\pm140$ & $4.3^{+1.4}_{-1.3}$ & $1.289\pm0.053$ & $1.43\pm0.06$ & & $1.54877430(120)$ & $1.769\pm0.064$ & $4.29\pm0.11$ & $6.044$ & $7.662$ & 18,20 \\
    WASP-121 & $6459\pm140$ & $1.5\pm1.0$ & $1.353^{+0.080}_{-0.079}$ & $1.458\pm0.030$ & & $1.27492550(25)$ & $1.183^{+0.064}_{-0.062}$ & $3.754^{+0.023}_{-0.028}$ & 6.005 & 7.929 & 21 \\
    WASP-122 & $5720\pm130$ & $5.11\pm0.80$ & $1.239\pm0.039$ & $1.52\pm0.03$ & & $1.71005328(14)$ & $1.284\pm0.032$ & $4.248\pm0.072$ & 6.147 & 7.895 & 18,22 \\
    WASP-167 & $7000\pm250$ & $1.29^{+0.36}_{-0.27}$ & $1.59\pm0.08$ & $1.79\pm0.05$ &  & $2.0219596(6)$ & $<8$ & $4.23^{+0.08}_{-0.07}$ & $5.528$ & $6.563$ & 18,24\\
    \hline
    \hline
    \enddata
    \tablecomments{\footnotesize \textbf{References:} 1. \cite{bonomo2017}, 2. \cite{bruno2016}, 3. \cite{penev2016}, 4. \cite{gaudi2017}, 5. \cite{oberst2017}, 6. \cite{chontos2019}, 7. \cite{collins2017qatar1}, 8. \cite{alsubai2017}, 9. \cite{ivshinawinn2022}, 10. \cite{wong2021}, 11. \cite{yee2023}, 12. \cite{torres2008}, 13. \cite{corteszuleta2020}, 14. \cite{esposito2017}, 15. \cite{ciceri2016}, 16. \cite{anderson2014}, 17. \cite{addison2016}, 18. \cite{kokori2023}, 19. \cite{gillon2014}, 20. \cite{barros2016}, 21. \cite{delrez2016}, 22. \cite{turner2016}, 23. \cite{stassun2017}, 24. \cite{temple2017}, 25. \cite{leonardi2024}.}
\end{deluxetable*}

As an example, \citet{siverd2012} adopted a stellar tidal quality factor of \(Q_\star' = 10^{8}\), an orbital period \(P_{\mathrm{orb}} \approx 1~\mathrm{day}\) for KELT-1, and found \(\tau_a = 3 \times 10^{4}\), yielding an orbital decay timescale $t_{\mathrm{decay}} \approx 0.3~\mathrm{Gyr}$. For the present study, to select candidate systems with potentially faster orbital decay (possibly due to IGWs, but not restricted to this mechanism), we fix \(Q_\star' = 10^{5}\) and \(P_{\mathrm{orb}} \approx 1~\mathrm{day}\). \par
We focus on systems whose decay timescales are < 500 Myr. To ensure unit consistency, the orbital period is converted from days to years, \jkt{$P_{\mathrm{orb}} = 1~\mathrm{day} = 1/{365.25}~\mathrm{yr}$,}{} leading to $\tau_a = 12\pi \times {t_{\mathrm{decay}}/}{Q^{'}_\star} \times 365.25$. Therefore, for the decay timescale range considered, \(\tau_a\) needs to satisfy 
\[
\log_{10} \tau_a  \lessapprox 7.
\]

Similarly, for the spin synchronization timescale, we arbitrarily take our timescale range as 0-500 Myr 
($t_{\mathrm{decay}} = t_{\mathrm{sync}}$). This gives 
$\tau_{\omega_\star} = 12\pi \tau_a$. 
Thus, we also focus on systems with
\[
\log_{10} \tau_{\omega_\star} \lessapprox 8.5.
\]

Based on these two criteria and a limit of $P_{\rm orb}\lesssim 2$~d, we first selected 19 HJ systems from the NASA Exoplanet Archive \citep{christiansen2025} and TEPCat \citep{southworth2011} (both accessed on 10 July 2023) as potential candidates for exhibiting orbital decay behavior within the suggested timescale ranges. Figure~\ref{fig:tau_comparison} presents the distribution of all selected systems within the corresponding parameter space.

In contrast to most of the selected systems, TrES-1 does not lie in the region of the shortest predicted orbital-decay timescales. However, TTV analyses \citep[e.g.][]{hagey2022,adams2024} indicate a significant period decrease for TrES-1\,b, suggesting that it remains a strong orbital-decay candidate despite its location in parameter space. Moreover, \citet{jackson2023} showed that the $\Delta\mathrm{BIC}$ in favor of an orbital decay model is expected to exceed the statistically decisive threshold ($\Delta\mathrm{BIC} > 50$) within the next few years. We therefore consider TrES-1\,b a noteworthy target for continued monitoring and include it in our sample, especially because the system has been observed for more than 20 years (Figure \ref{fig:ttv_plot_2}).

While KELT-9 stands out as an extreme case with its high temperature and stellar mass (see Table~\ref{tab:exoplanet_systems}), placing it outside the typical spectral range of our sample, it was retained because it falls within the relevant region of the parameter space for which we could plausibly detect orbital decay. In addition, previous studies \citep{ivshinawinn2022,harre2023} investigated the potential orbital period variation of KELT-9\,b, hypothesizing that it may exhibit signs of orbital decay; however, no statistically significant evidence supporting 
this was found.

Despite some individual peculiarities, TrES-1 and KELT-9 were included in our sample, yielding 20 HJ systems for further analysis. All parameters used throughout the study are listed in Table \ref{tab:exoplanet_systems}. Some systems meeting our selection criteria were excluded because reliable transit timing measurements or continuous coverage were unavailable for both space-borne and ground-based sources. These targets may be suitable for future studies with higher-cadence or longer-baseline monitoring, including HIP-65A b \citep{nielsen2020,Griffiths+25mn}, CoRoT-14 b \citep{tingley2011}, and HATS-70 b \citep{zhou2019}.

\section{Data and reduction}
\label{sec:Data}
\subsection{Observations}

\subsubsection{TUG100 Observations}

We used the T100 telescope at the Bak{\i}rl{\i}tepe Campus of the T\"urkiye National Observatories, over nine nights to observe four of our targets (see Table\,\ref{tab:observations}). The T100 was equipped with a SI 1100 CCD camera with a field of view (FoV) of 20$^\prime$$\times$20$^\prime$ sampled at 0.29$^{\prime\prime}$\,px$^{-1}$. We defocussed the telescope to improve the data quality \citep{southworth2009,basturk2015} and chose exposure times which gave the best signal-to-noise ratio (SNR) for each target.

\subsubsection{AUKR T80 Observations}

We observed seven targets during 10 nights with the T80 (Prof.\ Dr.\ Berahitdin Albayrak Telescope) at the Ankara University Kreiken Observatory. The CCD gave a FoV of 11.84$^\prime$$\times$11.84$^\prime$ at 0.69$^{\prime\prime}$\,px$^{-1}$. We used an SDSS $g'$, $r'$, $i'$ or $z'$ filter, a slight defocussing, and exposure times that gave $\approx$50 points during each transit.

\subsubsection{ATA050 Observations}

We observed a transit of Qatar-1 during one night using the ATA050 telescope, at Türkiye National Observatories in Erzurum, T{\"u}rkiye. The Apogee Alta U230 CCD camera gave a large FoV sampled at 2.77$^{\prime\prime}$\,px$^{-1}$. We used the Cousins $R$ band and an exposure time of 110\,s.

\subsubsection{IST60 Observations}

We observed three transits with the IST60 telescope at the Istanbul University Observatory (IUGUAM), at the \c{C}anakkale Onsekiz Mart University Ulup{\i}nar Observatory, T{\"u}rkiye. The Andor iXon Ultra 888 CCD camera had a FoV of 9.6$^\prime$~$\times$~9.6$^\prime$ at 2.56$^{\prime\prime}$\,px$^{-1}$.

\subsubsection{CDK20 and ODK20 Observations}
\label{sec:elsauce}

Two transits of HATS-18\,b were observed during 2 nights using a 50\,cm CDK telescope (CDK20 in Table\,\ref{tab:observations}), which is located at the El Sauce Observatory (UAI code X02) in Chile but remotely controlled from France. A luminance filter was used to maximize the SNR. Four transits of WASP-19 were observed using a 50\,cm ODK telescope (ODK20 in Table\,\ref{tab:observations}), a Moravian G4 16K CCD camera giving 0.53$^{\prime\prime}$\,px$^{-1}$, and an $R$ filter. The reduction of these data was performed using the Muniwin program from the photometry software package C-Munipack \citep{hroch2014}.

\subsubsection{Danish 1.54m Observations}
\label{subsec:danish}

Three transits of WASP-19\,b and one of WASP-103\,b were obtained using the 1.54\,m Danish Telescope at ESO La Silla, Chile, as a side-project of the MiNDSTEp microlensing observations \citep{Dominik+10an}. We used the Danish Faint Object Spectrograph and Camera (DFOSC), which has a 13.7$^\prime$$\times$13.7$^\prime$ field of view sampled at 0.39$^{\prime\prime}$\,px$^{-1}$. A Bessell $R$ or $I$ filter was used for each transit. The observations were obtained with the telescope defocussed. 

\begin{table*}
\centering
\caption{Summary of the observations.}
\label{tab:observations}
\resizebox{\textwidth}{!}{%
\begin{tabular}{lccccccccc}
\hline
\colhead{System} & \colhead{Date} & \colhead{Instrument} & \colhead{$N_{\rm f}$} &
\colhead{Filter} & \colhead{$T_{\rm exp}$} & \colhead{PNR} & \colhead{$\beta$} &
\colhead{Scatter} & \colhead{Airmass} \\
\colhead{} & \colhead{(yyyy-mm-dd)} & \colhead{} & \colhead{} &
\colhead{} & \colhead{(s)} & \colhead{(ppt)} & \colhead{} &
\colhead{(mmag)} & \colhead{}\\
\hline
HAT-P-23  & 2025-07-12 & T100  & 206 & V                   &  50 & 2.203 & 1.772 & 2.55 & 4.06 $\rightarrow$ 1.03 \\
HATS-18   & 2024-05-08 & CDK20 &  92 & L                   & 120 & 3.216 & 2.062 & 4.77 & 1.00 $\rightarrow$ 1.44 \\
HATS-18   & 2024-05-29 & CDK20 &  87 & L                   & 120 & 2.983 & 1.559 & 4.62 & 1.01 $\rightarrow$ 1.50 \\
Qatar-1   & 2014-08-17 & T100  &  64 & R                   & 120 & 0.678 & 1.554 & 1.00 & 1.23 $\rightarrow$ 1.13 \\
Qatar-1   & 2019-10-30 $^1$ & T100  & 161 & R                   &  90 & 1.071 & 2.767 & 1.35 & 1.14 $\rightarrow$ 1.79 \\
Qatar-1   & 2020-03-07 & T100  & 147 & R                   &  75 & 2.773 & 1.803 & 3.10 & 2.74 $\rightarrow$ 1.49 \\
Qatar-1   & 2020-07-03 & T100  & 110 & R                   & 100 & 1.453 & 1.370 & 1.91 & 1.45 $\rightarrow$ 1.14 \\
Qatar-1   & 2020-08-23 & IST60 & 104 & R                   & 120 & 4.172 & 1.876 & 5.55 & 1.17 $\rightarrow$ 1.68 \\
Qatar-1   & 2020-09-19 & IST60 &  84 & R                   & 110 & 3.699 & 1.239 & 4.89 & 1.31 $\rightarrow$ 1.91 \\
Qatar-1   & 2020-10-02 & T100  & 101 & R                   & 100 & 1.892 & 1.770 & 2.54 & 1.14 $\rightarrow$ 1.27 \\
Qatar-1   & 2020-10-09 & T100  &  70 & R                   & 120 & 1.122 & 1.332 & 1.61 & 1.26 $\rightarrow$ 1.75 \\
Qatar-1   & 2020-05-10 & ATA050 & 121 & R                   & 110 & 2.799 & 2.115 & 3.55 & 1.98 $\rightarrow$ 1.24 \\
Qatar-1   & 2025-03-10 & T80   & 202 & SDSS-$r'$           &  60 & 3.144 & 1.402 & 3.05 & 2.29 $\rightarrow$ 1.35 \\
Qatar-1   & 2025-06-09 & T80   & 126 & SDSS-$g'$           &  90 & 3.229 & 1.099 & 3.67 & 2.02 $\rightarrow$ 1.32 \\
Qatar-4   & 2022-11-01 & T80  & 125 & SDSS-$r'$           & 110 & 2.274 & 1.698 & 2.88 & 1.08 $\rightarrow$ 2.11 \\
TOI-2109  & 2023-05-01$^2$ & T80   &  81 & SDSS-$g'$           &   120  & 1.243 & 1.039 & 1.73 & 1.82 $\rightarrow$ 1.11 \\
TOI-2109  & 2024-06-27$^3$ & T100  & 204 & SDSS-$z'$           &  25 & 2.491 & 1.347 & 2.50 & 1.09 $\rightarrow$ 1.06 $\rightarrow$ 1.35 \\
TOI-2109  & 2024-07-01$^{1,2}$ & T80   & 133 & SDSS-$z'$           & 105 & 1.832 & 3.370 & 2.45 & 1.09 $\rightarrow$ 1.12 \\
TrES-1 & 2013-10-02 & Loiano 1.5m & 146 & SDSS-r' & 100 & 0.515& 1.802 & 6.74 & 1.00 $\rightarrow$ 1.96\\
WASP-3 & 2020-08-31 & IST60 & 81  & R & 75  & 2.860 & 2.184 & 4.66 & 1.01 $\rightarrow$ 1.59 \\
WASP-3 & 2022-07-06 & T80   & 220 & SDSS-$r'$ & 60  & 0.716 & 0.895 & 0.68 & 1.00 $\rightarrow$ 1.41 \\
WASP-3 & 2022-07-19 & T80   & 230 & SDSS-$r'$ & 60  & 1.167 & 1.859 & 1.12 & 1.01 $\rightarrow$ 1.29 \\
WASP-12   & 2025-02-02 & T100  & 107 & SDSS-$g'$           & 120 & 1.312 & 1.839 & 2.02 & 1.15 $\rightarrow$ 1.00 $\rightarrow$ 1.23 \\
WASP-12   & 2025-03-10 & T80   & 222 & SDSS-$i'$           &  75 & 1.526 & 1.745 & 1.59 & 1.08 $\rightarrow$ 2.45 \\
WASP-19   & 2013-02-15 & NTT &  55 & SDSS-$r'$ & 120 & 1.932 & 0.240 & 0.48 & 1.55 $\rightarrow$ 1.80 $\rightarrow$ 1.53 \\
WASP-19   & 2013-02-16 & NTT &  67 & SDSS-$r'$ & 120 & 2.220 & 0.278 & 0.55 & 1.68 $\rightarrow$ 1.05 \\
WASP-19   & 2013-02-18$^1$ & NTT &  63 & SDSS-$r'$ & 120 & 2.582 & 0.275 & 0.55 & 1.04 $\rightarrow$ 1.94 \\
WASP-19   & 2025-01-31$^3$ & ODK20 &  86 & R & 120 & 2.439 & 1.528 & 3.72 & 1.88 $\rightarrow$ 1.05 \\
WASP-19   & 2025-03-27 & ODK20 &  71 & R                   & 120 & 2.029 & 2.073 & 3.42 & 1.07 $\rightarrow$ 1.04 $\rightarrow$ 1.23 \\
WASP-19   & 2025-04-07 & ODK20 &  88 & R                   & 120 & 1.798 & 1.838 & 2.73 & 1.04 $\rightarrow$ 1.72 \\
WASP-19   & 2025-04-11$^3$ & ODK20 &  61 & R                   & 120 & 2.115 & 1.408 & 3.19 & 1.04 $\rightarrow$ 1.32 \\
WASP-19   & 2025-04-22 & Danish 1.54m &  259 & R & 30 & 1.692 & 0.884 & 1.33 & 1.09 $\rightarrow$ 1.96 \\
WASP-19   & 2025-05-07 & Danish 1.54m &  143 & I & 80 & 1.675 & 2.380 & 1.96 & 1.12 $\rightarrow$ 2.48 \\
WASP-19   & 2025-05-11 & Danish 1.54m &  323 & I & 20 &  2.164 & 1.651 & 1.49 &  1.08 $\rightarrow$ 1.65 \\
WASP-103  & 2022-07-04 & Danish 1.54m & 153 & R &  100 & 0.490 & 2.175 & 0.63 & 1.39 $\rightarrow$ 1.24 $\rightarrow$ 2.24 \\
WASP-103  & 2025-07-05 & T80   & 534 & SDSS-$g'$ &  25 & 1.689 &1.936  & 1.87 & 1.18 $\rightarrow$ 2.25 \\
WASP-114 & 2016-08-14 & CAHA & 163 & R$_{\rm c}$ & 120 & 1.944 & 0.789 & 1.09 & 1.30 $\rightarrow$ 1.12 $\rightarrow$ 1.98 \\
WASP-114  & 2020-08-26 & T80   & 111 & SDSS-$r'$/SDSS-$i'$ & 150 & 1.270 & 2.072 & 1.98 & 1.17 $\rightarrow$ 1.14 $\rightarrow$ 2.27 \\
WASP-121 & 2016-12-12 & GROND & 238 & SDSS-$g'$ & 40/50/60 & 0.581 & 1.677 & 6.28 & 1.73 $\rightarrow$ 1.01 $\rightarrow$ 1.04 \\
WASP-121 & 2016-12-12 & GROND & 238 & SDSS-$i'$ & 40/50/60 & 0.807 & 2.035 & 8.70 & 1.73 $\rightarrow$ 1.01 $\rightarrow$ 1.04 \\
WASP-121 & 2016-12-12 & GROND   & 238 & SDSS-$r'$ & 40/50/60 & 1.242 & 0.870 & 13.36 & 1.73 $\rightarrow$ 1.01 $\rightarrow$ 1.04 \\
WASP-121 & 2016-12-12 & GROND   & 238 & SDSS-$z'$ & 40/50/60 & 1.244 & 1.759 & 13.39 & 1.73 $\rightarrow$ 1.01 $\rightarrow$ 1.04 \\
\hline
\end{tabular}
}
\begin{flushleft}
{\footnotesize 
$^1$ Eliminated because its $\beta$-factor is larger than 2.5. \\
$^2$ Eliminated because its depth is more than 5$\sigma$ away from the average. \\
$^3$ Eliminated because it is an outlier on the TTV-diagram for both linear and quadratic models.
}
\end{flushleft}
\end{table*}

\subsubsection{GROND Observations}

We observed a transit of WASP-121 using the GROND (Gamma-Ray Burst Optical/Near-Infrared Detector; \citealt{greiner2008}) imager on the MPG 2.2\,m telescope at the ESO Observatory in La Silla (Chile). GROND obtains images simultaneously in four optical (SDSS $g^{\prime}$, $r^{\prime}$, $i^{\prime}$, $z^{\prime}$) and three infrared ($J$, $H$, $K$) bands, and has been used to observe many planetary transits previously \citep[e.g.,][]{mancini2014b,mancini2014a,mancini2016a,mancini2016b,southworth2015,southworth2017} including the characterisation of starspots \citep{mohler-fischer2013,mancini2013wasp19,biagiotti2024}. Our main aim in the current work is to measure a precise transit time, so we used only the optical bands. These have a FoV of $5.2^\prime \times 5.2^\prime$ sampled at $0.16^{\prime\prime}$\,px$^{-1}$ and were defocussed.

\subsubsection{CAHA Observations}
A transit of WASP-114 was observed with the Zeiss 1.23-m telescope at the Astronomical Observatory of Calar Alto in Spain, which has been extensively used for transit observations by our group \citep[e.g.][]{ciceri2013,mancini2017}. The DLR-MKIII CCD camera gave a FoV of $21.5^\prime$$\times$21.5$^\prime$ at $0.32^{\prime\prime}$\,px$^{-1}$. We used a Cousins $R$ filter, defocussing and CCD windowing to maximise the quality of the data. The sky was clear but not photometric.

\subsubsection{NTT Observations}
Three transits of WASP-19\,b were observed in February 2013 using the ESO New Technology Telescope (NTT), at La Silla, Chile, equipped with the ESO Faint Object Spectrograph and Camera (EFOSC2; \citealt{Buzzoni+84msngr}). We heavily defocussed the telescope and observed in the $R$ band (ESO filter \#784), obtaining photometric precisions of approximately 0.5\,mmag per point. Further details of this approach, and an example PSF can be found in \citet{Tregloan++13mn} and \citet{TregloanMe13mn}. The third transit contains a clear spot-crossing event.
\begin{figure*}
    \centering
    \includegraphics[width=\textwidth]{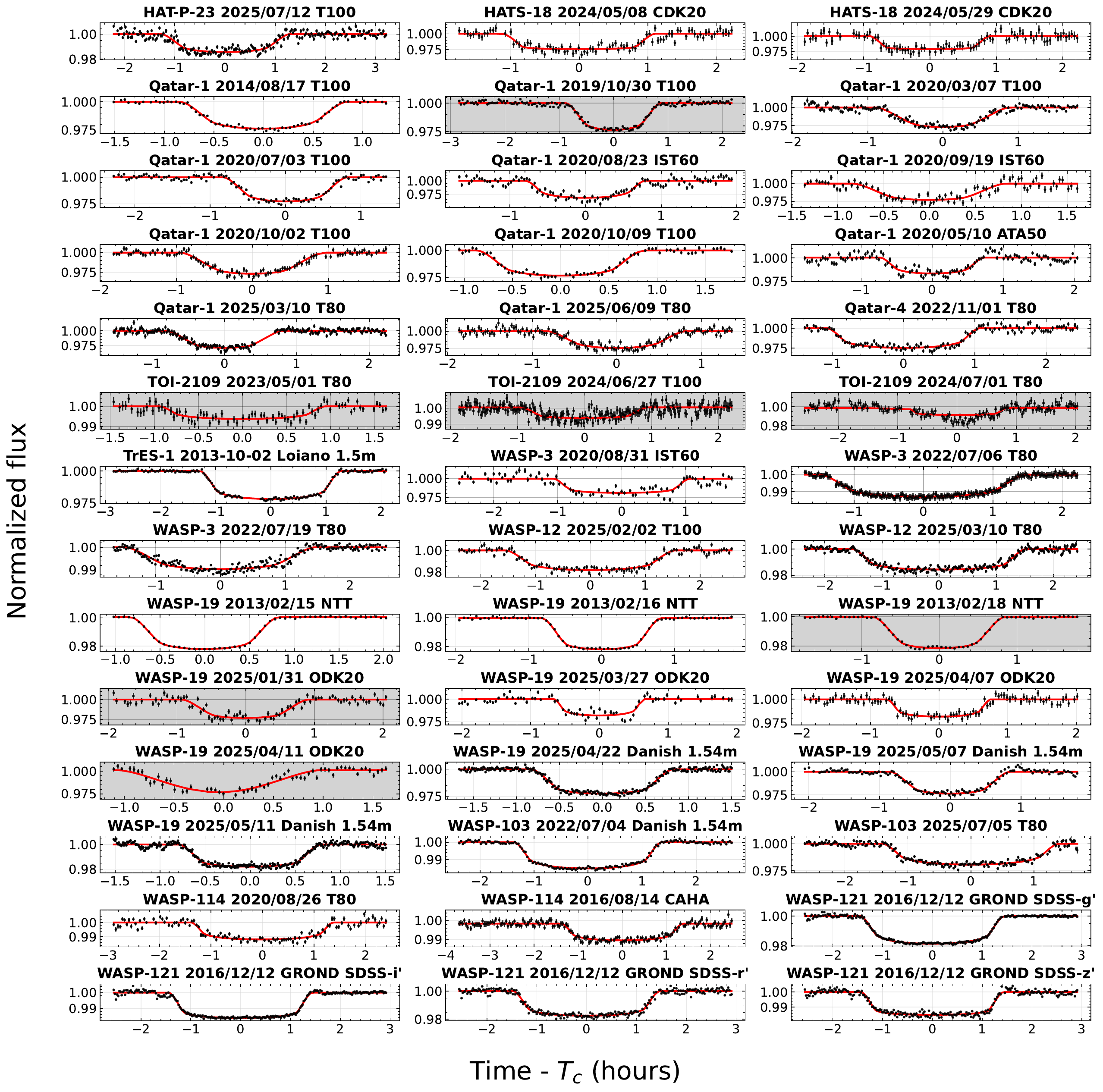}
    \caption{Light curves of the objects in our sample, as listed in Table \ref{tab:observations}. Individual exposures are shown in black, while the EXOFAST models are plotted in red. Light curves excluded from the sample due to PNR, $\beta$ or 3$\sigma$ clipping criterion are displayed with a grey background.}
    \label{fig:lc_plots}
\end{figure*}

\subsubsection{Loiano 1.52 m Observations}
\label{loiano}
A complete transit event of TrES-1\,b was recorded with the Cassini 1.52\,m telescope from the Loiano Observatory (Italy), which at that time was part of the Astronomical Observatory of Bologna. The BFOSC (Bologna Faint Object Spectrograph and Camera) instrument was used to give images with a FoV of 13.0$^\prime$$\times$12.6$^\prime$ at 0.58$^{\prime\prime}$\,px$^{-1}$. The CCD was windowed, a Gunn-$i$ filter was used, and the telescope was defocussed.

\subsection{Data Reduction}

\subsubsection{AstroImageJ}
All the observations performed with the T80, T100, IST60 and ATA050 telescopes were reduced using the AstroImageJ (hereafter AIJ; \citeauthor{astroimage2017} \citeyear{astroimage2017}) software package, where standard calibration procedures (bias, dark, and flat corrections) were applied with the corresponding calibration frames obtained during the same nights. \jkt{When required, 2×2 pixel binning was adopted to reduce exposure and readout times.}{}The observation times were converted to barycentric Julian Dates (hereafter BJD$_{\rm TDB}$) before performing ensemble photometry \citep{honeycutt1992} with AIJ, using a set of comparison stars selected according to their brightness, color, and photometric stability in the observed passband. The aperture centers were manually determined to avoid misidentifications caused by centroid algorithms, especially in cases where defocused images produced \jkt{“ring-like” stellar profiles or collimated}{complex} PSFs. Finally, corrections for airmass and sky/pixel ratio were applied, and the relative fluxes derived with AIJ were normalized by fitting a line to the out-of-transit data points.

\subsubsection{TESS}
The Transiting Exoplanet Survey Satellite (TESS) observed all objects in our sample \jkt{for at least one sector, with the complete list of sectors provided in Table \ref{tab:tess_sectors}}{except WASP-103; see Table \ref{tab:tess_sectors}}. For \jkt{all these sectors,}{each sector} we downloaded the 2-min cadence mode products of the Science Processing Operations Center (SPOC) pipeline \citep{jenkins2016} from the Barbara A. Mikulski Archive for Space Telescope (MAST) of the Space Telescope Science Institute (STScI) of NASA\footnote{ https://mast.stsci.edu/portal/Mashup/Clients/Mast/Portal.html}. From the SPOC pipeline, we obtained the data validation timeseries (DVT) light curve products utilizing \jkt{}{the} {\sc lightkurve} package \citep{lightkurve2018} since instrumental artifacts and stellar variabilities were corrected in the {\sc lc\_detrend} column of the relevant DVT-files. To improve accuracy and precision, we discarded apparent outliers linked to instrumental effects and split the full data set into smaller parts focused on single transit events using the ephemeris information provided in the corresponding DVT-files. \par

\subsubsection{\textsc{defot}}

The data from the Danish telescope, the NTT, GROND, CAHA and Loiano were reduced using the {\sc defot} pipeline \citep{Me+09mn2,southworth2014}. This utilizes the {\sc idl}\footnote{{\tt https://www.ittvis.com/idl/}} implementation of the {\sc aper} routine from {\sc daophot} \citep{Stetson87pasp} contained in the NASA {\sc astrolib} library\footnote{{\tt http://idlastro.gsfc.nasa.gov/}} to perform aperture photometry. We calculated master bias and flat-field images but did not use them as they increased the scatter in the data without changing the transit shapes.

A differential-magnitude light curve was generated for each transit observation by constructing a composite comparison star. This star was made by iteratively adjusting the weights of individual stars and the coefficients of a low-order polynomial to minimise the scatter in the data outside transit. The timestamps for the midpoint of each image were taken from the headers of the {\sc fits} files and converted to BJD$_{\rm TDB}$ using routines from \citet{Eastman++10pasp}.

\subsubsection{Exoplanet Transit Database and Literature work}

Additional photometric data were obtained from the Exoplanet Transit Database\footnote{\tt https://var.astro.cz/en/Home/ETD} (hereafter ETD; \citealt{poddany2010}) and previous studies. For ETD light curves, we only included light curves with a data quality factor of 2 or lower (DQ $\leq$ 2) to avoid the influence of gaps or high noise levels, which can interfere with accurate mid-transit time measurements. Airmass values were first calculated for the observing site following the formulation given in \citet{hiltner1962}, and all observation times provided in JD$_{\rm UTC}$ or HJD$_{\rm UTC}$ were converted to BJD$_{\rm TDB}$ using the relevant {\sc{astropy}} constants and functions \citep{astropy2013,astropy2018}. Airmass detrending and out-of-transit normalization were subsequently performed in AIJ. Literature data were processed in a similar manner: time stamps were converted to BJD$_{\rm TDB}$ when necessary, and light curves that were not already normalized or airmass-detrended underwent the same AIJ procedures to ensure a homogeneous data set.

\subsection{Light Curve Selection Criteria}
\label{subsec:selection_criteria}
Before performing the construction of the TTV diagram, we excluded light curves that contained significant gaps within the transit or showed strong correlated noise, particularly during ingress and egress. The remaining light curves were modelled with {\sc exofast} \citep{eastman2013} (see \S~\ref{subsec:modeling_lc} for details), after which we computed the photometric noise rate (PNR; \citealt{fulton2011}) from the residuals as a measure of white noise. Light curves with PNR values exceeding the transit depth (i.e. PNR $> \delta$) were discarded. To assess red noise, we binned the residuals over intervals of ingress/egress duration $\pm5$ minutes in 1-minute steps and calculated the $\beta$ values following \citet{winn2008}. Light curves with a median $\beta$ greater than 2.5 were rejected. Additionally, we excluded any light curves where the transit depth was a $5\sigma$ outlier for the corresponding planet. Finally, we omitted light curves that deviated from both the linear and quadratic fits to the TTV diagram by more than $5\sigma$. \par
The only exception in our sample is KELT-9, for which the PNR--$\beta$ analysis is not adequate. This system is strongly affected by gravity darkening due to the rapid stellar rotation ($v\sin{i} = 111.4 \pm 1.3$ km s$^{-1}$; \citeauthor{gaudi2017} \citeyear{gaudi2017}), which makes the star oblate and produces a temperature gradient between the hotter poles and the cooler equator. As a result, transits across different stellar latitudes yield asymmetric light-curve shapes that cannot be well described by standard models, leading to strongly correlated residuals \citep{ahlers2020,harre2023}. In addition, stellar pulsations with a characteristic period of about 7.59 hours have been reported in KELT-9 \citep{wong2019}, introducing quasi-periodic variability on time scales comparable to the transit duration ($T_{14} \approx 4$ hrs; \citeauthor{gaudi2017} \citeyear{gaudi2017}). These two effects together generate excessive levels of time-correlated (red) noise, which results in artificially high $\beta$ values and the rejection of the majority of light curves if the same criteria were applied. For this reason, KELT-9 data were treated as a special case, and its light curves were included in the analysis without the PNR--$\beta$ filtering.\par

\section{Analysis and Results}
\label{sec:analysis_results}
\subsection{Light-curve modeling and measurements of mid-transit times}
\label{subsec:modeling_lc}

We adopted the procedure described in \citet{basturk2022} for modeling the light curves and determining the mid-transit times and their uncertainties. In summary, the light curves were modelled with {\sc exofast} \citep{eastman2013}, after converting the observation times into BJD$_{\rm TDB}$ and detrending them for airmass effects using AIJ when necessary. Throughout the study, we used the web version of the code ({\sc exofast-v1})\footnote{https://exoplanetarchive.ipac.caltech.edu/cgi-bin/ExoFAST/nph-exofast}. The values of the stellar atmospheric parameters were \jkt{automatically}{}taken from the NASA Exoplanet Archive and set as Gaussian priors, while the orbital periods of the planets were fixed. The limb-darkening coefficients were assigned uniform priors, retrieved from \citet{claret2011} based on the stellar parameters and the observational passbands. For passbands not directly available, the closest match in transmission curve was adopted (e.g., the $I$ band for the TESS passband and \emph{CoRoT} for Clear-filter observations). Since EXOFAST does not account for spot-induced asymmetries in transit light curves, spot-crossing features were neither explicitly modelled nor masked, and the light curves were fitted assuming a symmetric transit profile \citep{plavchan2020}. The mid-transit time, initially treated as a free parameter, was derived from the contact points of the light curve model. The program outputs this value along with its associated uncertainty and other fitted parameters—such as the total and full transit durations and the transit depth—which were subsequently used to verify the internal consistency of the results and their agreement with the known system parameters. \par

\begin{table}
\centering
\caption{Measured mid-transit times and their uncertainties.}
\setlength{\tabcolsep}{4pt}
\renewcommand{\arraystretch}{1.2}
\begin{tabular}{lcc}
\hline
System & $T_0$  & $\sigma$ \\
& (BJD$_{\rm TDB}$) & (days) \\
\hline
HAT-P-23 & 2460869.39168 & 0.00070 \\
HATS-18	 & 2460439.60668 &	0.00082\\
HATS-18 & 2460460.55349 & 0.00086 \\ 
Qatar-1	& 2456887.31479 & 0.00019 \\
Qatar-1	& 2456234.10349 & 0.00023 \\
... & ... & ... \\
\hline
\end{tabular}
\begin{flushleft}
    {\footnotesize \textbf{Note.} This table represents the form of timing data; the full table is available at the CDS. The order of measured mid-transit times is as the same as Table \ref{tab:observations}.}
\end{flushleft}
\label{tab:mid_transit_times}
\end{table}

\subsection{Transit Timing Variations (TTV) Analysis}
\label{sec:ttv_analysis}
After selecting the light curves based on the selection criteria described in \S~\ref{sec:target_selection}, we constructed our TTV diagrams for each system in the sample according to the reference ephemeris information given in Table \ref{tab:ref_ephemeris}. For each target, we took into account all transit times from this study, along with those reported in the literature that satisfied the following conditions: ($i$) the time system was explicitly specified as BJD$_{\rm{TDB}}$, ($ii$) the light curve consists of both the ingress and egress of the transit and ($iii$) we were not able to access its light curve to model ourselves. In the end, we \jkt{opted}{obtained} a total of 2930 minimum times for the whole sample, as listed in Table \ref{tab:lcs_after_pnr_beta}. \par
We fitted two models to the TTV diagram. The first model assumes a constant orbital period over time
\begin{equation}
    T(E) = T_0 + P_{\rm orb} \times E .
\label{eq:lineer}
\end{equation}
Here, $T_0$ is the reference time of minimum, which is also given in Table \ref{tab:ref_ephemeris}. \par
On the other hand, the second model assumes a secular variation in the orbital period, allowing it to change linearly with time. In this case, the ephemeris can be expressed as
\begin{equation}
T(E) = T_0 + P_{\rm orb} \times E + \frac{1}{2}\frac{dP_{\rm orb}}{dE} \times E^2,
\label{eq:quadratic}
\end{equation}
where ${dP_{\rm orb}}/{dE}$ represents the time derivative of the orbital period. In the case of exoplanetary systems, the measured period variations are typically on the order of a few nano-days per orbital cycle (e.g., WASP-12\,b,~  $dP_{\rm orb}/dE = -5.364 \times 10^{-10}$~d~cycle$^{-1}$; \citeauthor{kutluay2023}, \citeyear{kutluay2023}). So, to help the reader along the way, it is more beneficial to read the results in a more meaningful scale. A positive value of the quadratic term indicates that the orbital period is increasing over time, whereas a negative value implies a decreasing period.

In both cases, we have drawn samples from Gaussian priors for the model parameters, with the priors centered on values obtained from an initial non-linear least-squares fit and widths set according to the corresponding uncertainties. Sampling from these prior distributions was performed using 16 Markov chains of 5000 iterations, 500 of which are discarded as burn-in. For each sample, the log-likelihood was computed, allowing the construction of posterior distributions for the model parameters. The posterior samples were subsequently obtained after discarding an initial burn-in period. The probabilistic fitting was conducted using the current (fifth) version of {\sc pymc5} \footnote{https://www.pymc.io/welcome.html} \citep{pymc2023}, while the preliminary non-linear least-squares fits were performed with {\sc lmfit} \citep{newville2016} to obtain reliable initial values for the fit parameters. \par
To assess which model provides a better representation of the timing data, we employed the reduced chi-square statistic ($\chi_\nu^2$), the Bayesian Information Criterion (BIC; \citeauthor{schwarz1978}~\citeyear{schwarz1978}; \citeauthor{liddle2007}~\citeyear{liddle2007}) and the Akaike Information Criterion (AIC; \citeauthor{akaike1974}~\citeyear{akaike1974}). To further quantify the relative performance between the linear and quadratic models, we computed the difference in their Information Criterion values, defined as
\begin{equation}
\label{eq:bic}
\Delta \mathrm{BIC} = \mathrm{BIC}_{\mathrm{quadratic}} - \mathrm{BIC}_{\mathrm{linear}},
\end{equation}
and
\begin{equation}
\Delta \mathrm{AIC} = \mathrm{AIC}_{\mathrm{quadratic}} - \mathrm{AIC}_{\mathrm{linear}}.
\end{equation}

\begin{figure*}
    \centering
    \includegraphics[width=\textwidth]{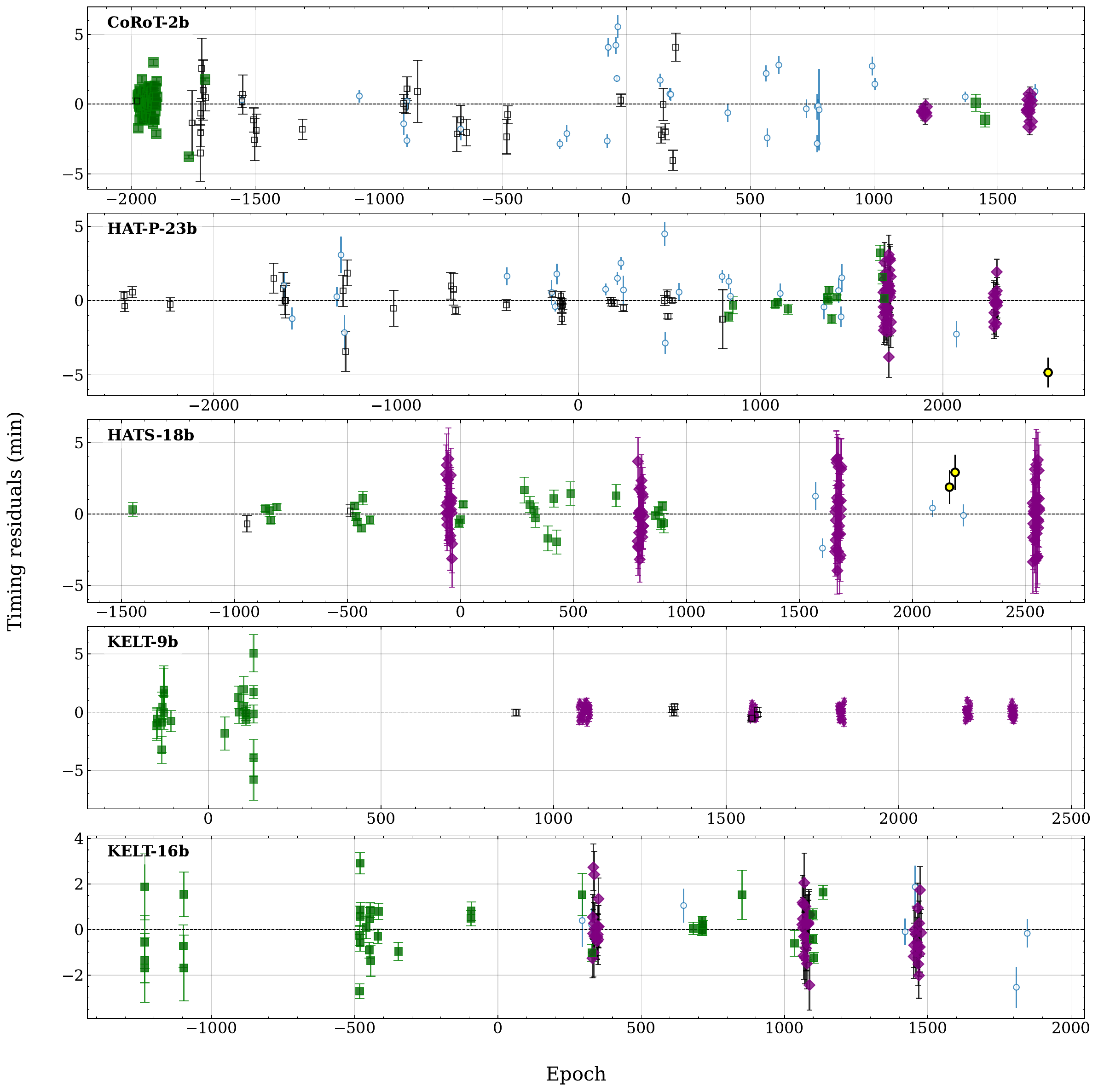}
    \caption{Linear residuals of the TTV diagrams for CoRoT-2\,b, HAT-P-23\,b, HATS-18\,b, KELT-9\,b and KELT-16\,b based on observations from various sources: ETD data (blue empty circles), obtained and analyzed literature light curves (green squares), adopted literature transit times (empty black squares), this work (black and yellow circles) and TESS observations (purple circles).}
    \label{fig:ttv_plot_1}
\end{figure*}

\begin{figure*}
    \centering
    \includegraphics[width=\textwidth]{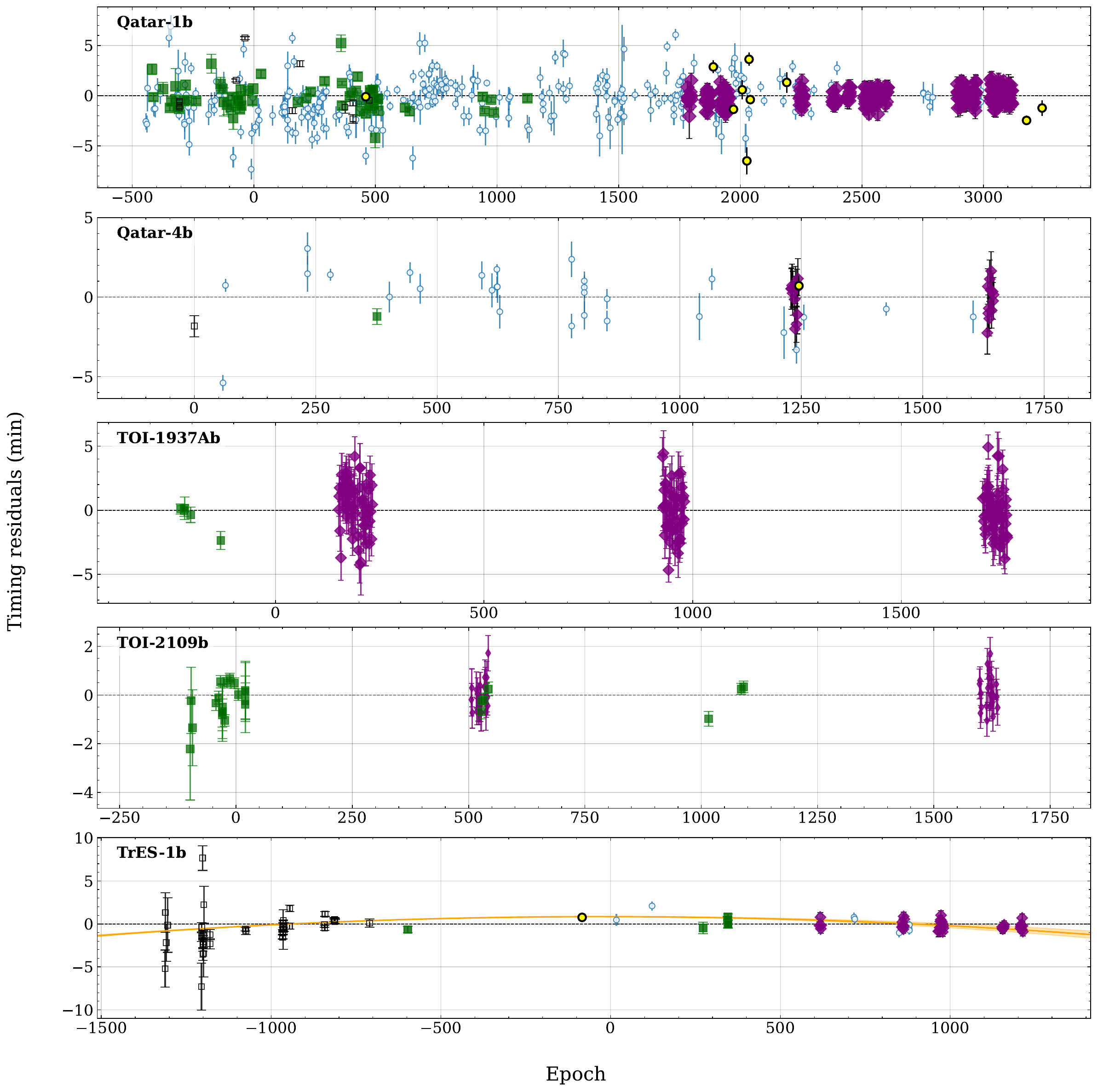}
    \caption{Same as Figure \ref{fig:ttv_plot_1}, but for Qatar-1\,b, Qatar-4\,b, TOI-1937A\,b, TOI-2109\,b and TrES-1\,b. Since $|\Delta {\rm BIC}| > 10$ for TrES-1\,b case, we also present the median orbital decay model represented with the orange line and the shaded band indicating its $3\sigma$ uncertainty range.}
    \label{fig:ttv_plot_2}
\end{figure*}

\begin{figure*}
    \centering
    \includegraphics[width=\textwidth]{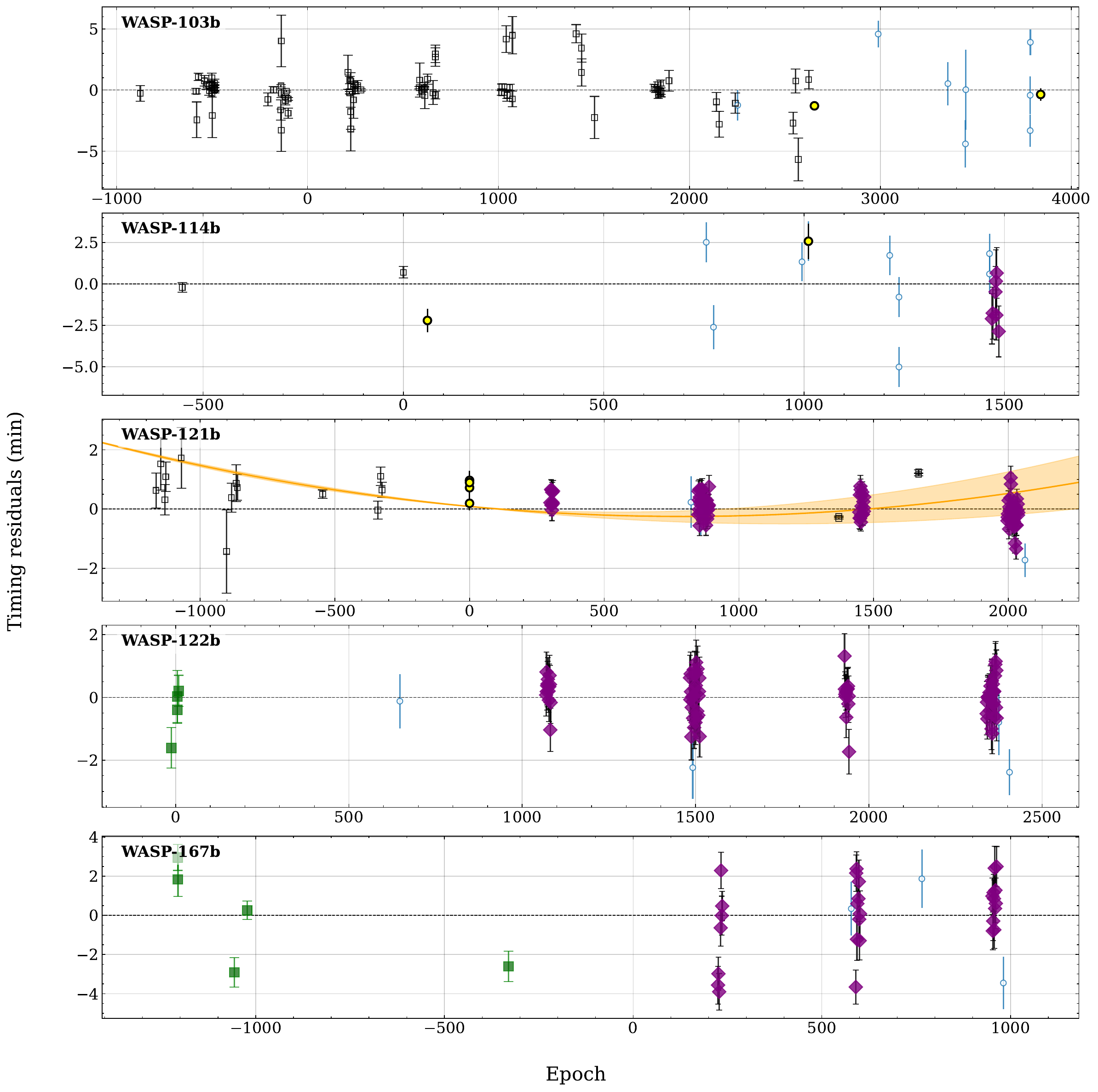}
    \caption{Same as Figure \ref{fig:ttv_plot_1}, but for WASP-103\,b, WASP-114\,b, WASP-121\,b, WASP-122\,b and WASP-167\,b. Since $|\Delta {\rm BIC}| > 10$ for the WASP-121\,b case, we also present the median quadratic model represented with the orange line and the shaded band indicating its $3\sigma$ uncertainty range.}
    \label{fig:ttv_plot_4}
\end{figure*}

Following the conventional interpretation \citep{raftery1995}, a difference of $|\Delta \mathrm{BIC}| > 10$ is considered strong evidence in favor of the model with the lower BIC value and likewise for $|\Delta \mathrm{AIC}|$. For systems with $|\Delta \mathrm{BIC}| < 10$, the improvement of the leading model is not statistically significant. In such cases, we only updated the linear ephemeris using Eq.~\ref{eq:lineer}. Figures \ref{fig:ttv_plot_1}, \ref{fig:ttv_plot_2}, \ref{fig:ttv_plot_3} and \ref{fig:ttv_plot_4} present the TTV diagrams of the systems with updated ephemerides. We investigated whether periodic or quasi-periodic variations are present in the TTV data based on Lomb-Scargle (hereafter LS) periodograms \citep{lomb1976,scargle1982} by employing the relevant function in the {\sc astropy} package \citep{VanderPlas2018}. We only considered the cases where we observed peaks with a False Alarm Probability (FAP) value smaller than 0.1\% and investigated the potential reasons of the signal.

\subsection{Modeling the efficiency of tidal dissipation using {\sc MESA}}
\label{stellarmodels}

To complement the TTV analysis, we have also constructed models of each of the host stars in our sample based on the parameters reported in Table~\ref{tab:exoplanet_systems} using {\sc MESA} version r24.08.1 \citep{Paxton2011,Paxton2013,Paxton2015,Paxton2018,Paxton2019,Jermyn2023}, along with stellar parameters from MIST \citep{MIST02016,choi2016}. In each case, we adopted an initial metallicity of 0.02 (unless otherwise specified) and the shortest reported rotation period constraint in the literature (purely for computation of the tidal period -- the stellar models themselves omit rotation). We integrated these models until the end of the main sequence or to the maximum reported age in Table~\ref{tab:exoplanet_systems}, whichever is shorter, with a few exceptions that integrated for longer. This allowed us to obtain radial profiles ($r$ is the spherical radius from the stellar center) for the stellar density $\rho(r)$, pressure $p(r)$, gravitational acceleration $g(r)$, and other variables, as a function of stellar age. Using these profiles, we calculated the tidal response following the approach described in \citet{barker2020}. We first computed the equilibrium tide response, and hence the viscous dissipation of this component (see Eq.~23 in \citeauthor{barker2020}~ \citeyear{barker2020}). We then computed the wavelike response and the resulting dissipation using Eqs.~\ref{eq:IW} and \ref{eq:IGW}.

\subsection{Individual results}

\subsubsection{CoRoT-2}

CoRoT-2\,b (CoRoT-Exo-2\,b) is the second transiting planet discovered by the \emph{Convection, Rotation and planetary Transits} (CoRoT) mission \citep{alonso2008}. 
Several studies have investigated possible orbital period changes driven by tides. While early and recent analyses have yielded conflicting results, some finding no significant decay and others reporting evidence in favor of orbital decay, the system remains an important and debated target for tidal evolution studies \citep{ozturkerdem2019,ivshinawinn2022,wang2024,adams2024}.
Before the TTV analysis results, it should be noted that the light curve data originally provided by  \citet{ozturkerdem2019} were obtained from the corrected versions of {\protect \citet{adams2024}}, who pointed out that the time stamps in \citet{ozturkerdem2019} were given in JD rather than HJD or BJD.

The obtained values of $\Delta \text{AIC} = 1.93$ and $\Delta \text{BIC} = 4.97$ indicate only a mild preference for the linear model, but this preference is inconclusive. The quadratic model yields a positive period change of $3.4 \pm 1.9~\mathrm{ms~yr^{-1}}$, which is unexpected from tidal theory, since for $P_\mathrm{orb} < P_\mathrm{rot}$ we predict angular momentum transfer from the orbit to the stellar rotation. We therefore constrain the stellar tidal quality factor \( Q_\star^{\prime} \) using the 3$\sigma$ lower limit of the quadratic coefficient, \( a^{lim}_{1} = 0.5 \times dP/dE = -1.68 \times 10^{-11} \)~d~cycle\(^{-1}\), following \citet{southworth2022}. This corresponds to a 99.7\% confidence limit of \( Q_\star^{\prime} > (5.27 \pm 0.87) \times 10^{5} \).

The strongest peak in the LS periodogram of the TTV diagram for CoRoT-2\,b is observed at 470.0~d with a FAP value of $3.2 \times 10^{-5}$. However, when we phase the TTV diagram with this periodicity and the integer divisors of it, we do not observe any agreement with the periodic models. Although there is a nearby, potentially bound source (Gaia DR3 4287820852697823872) in Gaia data, this $\rm G=15^m.46$-star is unlikely to be the source of the signal because it is separated from CoRoT-2 by $4.08 \pm 0.03$ arcsec at $\sim210$ parsecs, which corresponds to a linear distance more than $850$ au. There have also been no mentions of any other outside perturber in the literature.

Theoretical models of CoRoT-2 following the approach outlined in \S~\ref{stellarmodels} provide us with $Q^\prime_\mathrm{IGW} = 1.94\times 10^6$ at an age of 2.7 Gyr (the value is not strongly sensitive to age), assuming fully damped gravity waves. The star is of solar type with a radiative core, and the critical planetary mass required for wave breaking is $M_\mathrm{crit}/M_{\rm J}=1.4$ at the same age. Hence, the planet is predicted to be sufficiently massive to induce wave breaking in the stellar core, and we predict the planet to be spiralling into the star, but at a slower rate (larger $Q_\star^{\prime}$) than the lower bound observational constraint. Equilibrium tides are predicted to be negligibly weak, and inertial waves cannot be excited in the convective envelope (for an aligned orbit) as the tidal period exceeds $P_\mathrm{rot}/2$.


\subsubsection{HAT-P-23}

HAT-P-23\,b is a HJ with a mass of \(1.97~M_{\mathrm{J}}\) and an inflated radius of \(1.37~R_{\mathrm{J}}\), orbiting a G0-type star of \(\sim 1.1~M_{\odot}\) with \(P_{\mathrm{orb}} = 1.21~\mathrm{days}\) \citep{Bakos2011,ciceri2015,bonomo2017}. The orbit is approximately aligned with the stellar spin axis (\(\lambda = 15^{\circ} \pm 22^{\circ}\); \citeauthor{moutou2011} \citeyear{moutou2011}). Several studies have searched for evidence of orbital decay and constrained the stellar tidal quality factor $Q_\star^{\prime}$, consistently finding only lower limits and no significant evidence for ongoing decay. The most recent analysis likewise concludes that orbital decay has not yet been detected in the HAT-P-23 system \citep{maciejewski2018,patra2020,maciejewski2022,alvarado2024}.

As in previous works, our analysis reveals no significant evidence of orbital decay for this planet. We found $\Delta \text{AIC} = 1.77$ and $\Delta \text{BIC} = 4.60$, favoring the constant period model over the changing orbital period model. The median quadratic fit suggests a period change of $0.80 \pm 0.69 $ ms yr$^{-1}$. Considering the spin-orbit relationship, this outward migration would not be predicted theoretically. Still, we calculated a lower limit of the stellar tidal quality factor $Q_\star^{\prime} > (1.41 \pm 0.45) \times 10^5$ using the 99.7\% level confidence level, with \( a^{lim}_{1} = -4.19 \times 10^{-11} \)~d~cycle\(^{-1}\).

Theoretical models of HAT-P-23 indicate that fully damped gravity waves would provide $Q^\prime_{\rm IGW}=6.7\times 10^5$ at 4 Gyr. The star possesses a convective core, and wave breaking of these waves near the center is not predicted, which makes it difficult to justify the fully damped regime being relevant \citep[unless magnetic wave conversion is in operation, which is not predicted according to][]{duguid2024}. Hence, we would not clearly predict orbital decay, though if the gravity waves are fully damped, they would provide such a value of $Q^\prime_{\rm IGW}$. The star likely rotates sufficiently slowly ($P_{\rm rot} = {7.015}$~d; \citeauthor{salisbury2021}~\citeyear{salisbury2021}) to preclude inertial wave excitation.

\subsubsection{HATS-18}

HATS-18\,b ($M_{\rm p} =1.98~M_{\rm J}$, $R_{\rm p} = 1.34~R_{\rm J}$), orbiting a G\,V type solar-like star with $P_{\rm orb} \approx 0.83$~days, was discovered by \citet{penev2016}. The nearly solar-mass ($1.04~M_\odot$) and solar-age ($4.2 \pm 2.2$~Gyr) host star would be expected to rotate with a period of $\sim 30$~days \citep{barnes2016}, yet observations indicate a much faster rotation of $\sim 9$~days, possibly due to tidal spin-up by the close-in giant planet. Several studies have investigated orbital decay; no detectable period change has been found and the stellar tidal quality factor is constrained to $Q_\star' > 1.3 \times 10^5$ \citep{southworth2022} and refined to $Q_\star' > 3.5 \times 10^5$ with recent TESS observations \citep{maciejewski2024}.

Overall, our analysis yields $\Delta \text{AIC} = 1.72$ and $\Delta \text{BIC} = 4.65$, suggesting that there is no evidence to favor orbital decay within the system. 
Hence, similar to the cases of CoRoT-2\,b and HAT-P-23\,b, we constrained the stellar tidal quality factor $Q_\star^{\prime}$ using the $3\sigma$ lower limit $a^{lim}_{1} = -1.364 \times 10^{-10} ~\text{d}\,\text{cycle}^{-1}$. This provides the constraint $Q_\star^{\prime} > 6.8 \pm 2.0 \times 10^5$, which is in agreement with the findings of \cite{southworth2022} and \cite{maciejewski2024}.

Based on the models presented in \cite{southworth2022}, we predict $Q^\prime_\mathrm{IGW}\approx 1.2\times 10^5$ for gravity waves in the fully damped regime. This regime is likely to be appropriate because the planet's mass exceeds the critical mass required for wave breaking at its current age. Hence, we would predict this planet to be undergoing orbital tidally-driven orbital decay. This is an exciting system for future studies, as we predict orbital decay with a value of $Q^\prime_\mathrm{IGW}$ that is quite similar to the current observational constraint for its lower bound. Hence, future observations of this system would be worthwhile to constrain tidal theory.

\subsubsection{KELT-9}

This system lies at the extreme end of the sample, particularly in terms of stellar temperature. KELT-9\,b, the hottest known exoplanet with $T_{\rm eq} = 4050$~K, is a $2.88~M_{\rm J}$ inflated HJ ($1.9~R_{\rm J}$) orbiting a B9.5--A0 type star of $1.98~M_\odot$ and $T_{\rm eff} \sim 10{,}000$~K with $P_{\rm orb} \approx 1.5$~days \citep{gaudi2017,hoeijmakers2019}. The host is a fast rotator ($v\sin i = 111.4 \pm 1.3~\text{km s}^{-1}$ or $P_{\rm rot} = 18.96 \pm 0.34$~h; \citeauthor{jones2022} \citeyear{jones2022}), with $P_{\rm orb} > P_{\rm rot}$, implying tidal angular momentum transfer from the star to the orbit and possible orbital expansion. \citet{harre2023} found no evidence for a secular orbital period change and attributed the TTV signal to apsidal precession, despite the orbit being reported as circular by \citet{gaudi2017}. This scenario is supported by RV data from \citet{stephan2022} and is discussed further in \S~\ref{subsec:apsidalmotion}.

Our analysis resulted in $\Delta \text{AIC} = 1.97$ and $\Delta \text{BIC} = 4.75$, indicating that there is no statistically significant evidence for orbital decay or expansion. The orbital decay model suggests a change of $0.45\pm 0.96 ~\text{ms yr}^{-1}$. To constrain $Q_\star^{\prime}$, we took the $3\sigma$ lower limit $a^{lim}_{1} = -6.751 \times 10^{-11} ~\text{d}\,\text{cycle}^{-1}$. However, since the HJ has a retrograde and polar orbit, we took the tidal factor\footnote{The tidal factor $f$ corresponds to the numerical coefficient $-27/2$ in Eq.~\ref{eq:modified_tidal_quality_factor}, which arises from the equilibrium tide formalism under the assumption of a prograde orbit.} as $-135/16$ rather than $-27/2$ (see Table 2 in \citeauthor{harre2023} \citeyear{harre2023}) while utilizing Eq. \ref{eq:modified_tidal_quality_factor}. Thus, we provide a lower bound to $Q_\star^{\prime} > (1.21 \pm 0.37) \times 10^6$.

Our theoretical models of KELT-9 provide $Q^{'}_{\rm IGW}>10^{12}$ for the $l=m=2$ tide at 0.3 Gyr based on waves excited from the interface with the convective envelope, and hence negligible tidal migration. On the other hand, the star rotates rapidly enough for inertial waves to be excited in the envelope, leading to $Q^\prime_{\rm IW}\sim 10^{10}$ for a similar age. Equilibrium tides are also predicted to be weak with $Q^\prime_\mathrm{eq}\sim 3\times 10^{12}$. Gravity waves launched from the convective core may be more important in this star but they are unlikely to provide sufficient dissipation to predict observable orbital expansion. On the other hand, the polar orbit suggests other tidal components could be excited than the ones we consider here.


\subsubsection{KELT-16}
\label{subsub:kelt16}

KELT-16\,b ($M_{\rm p}=2.75~M_{\rm J}$, $R_{\rm p}=1.415~R_{\rm J}$) is one of the few planets with $P_{\rm orb}<1$~day and is therefore a promising target for detecting orbital period changes \citep{oberst2017}. It orbits an F7V star ($1.21~M_\odot$, $1.36~R_\odot$) that has a distant M-dwarf companion at $\sim 300$~au, which may have driven the planet inward via Kozai--Lidov oscillations \citep{oberst2017}. With $T_{\rm eq}=2453$~K, KELT-16\,b is classified as an ultra-hot Jupiter. Using tidal evolution models, \citet{oberst2017} showed that the planet could be tidally disrupted for $Q^{'}_\star \sim 10^5$.
Several studies have searched for evidence of orbital decay. While early analyses reported constraints on $Q^{'}_\star$ and tentative period variations \citep{maciejewski2018,patra2020}, subsequent investigations have found no statistically significant evidence for ongoing orbital decay to date \citep{maciejewski2022,mancini2022,harre2023,alvarado2024}.

From our analysis, we found $\Delta {\rm AIC} = -0.68$ and $\Delta {\rm BIC} = 1.88$. Hence, we still do not have convincing evidence from TTV supporting the shrinking orbit of KELT-16\,b. Our quadratic fit suggests orbital decay, though with an orbital period change of $-16.8 \pm 4.5 ~{\rm ms~yr}^{-1}$. Using the 95\% confidence upper limit on the decay rate, we thus constrain the stellar tidal quality factor to $Q_\star^{\prime} > (1.03 \pm 0.21) \times10^6$.\par
Theoretical models of KELT-16\,b predict $Q^\prime_\mathrm{IGW}\approx 1.5\times 10^6$ at 2 Gyr, albeit a slightly younger age than the one listed in Table \ref{tab:exoplanet_systems} \citep{oberst2017}. More efficient dissipation would be predicted for later ages. The star possesses a convective core and wave breaking of these waves is not predicted. Magnetic wave conversion \citep{duguid2024} remains a possibility to justify the fully damped gravity wave regime, but this is uncertain, depending on the age of the star. The star probably rotates too slowly for inertial waves to be excited. 
Our theoretical predictions are close to the current observational constraint, making this an exciting system for follow-up studies.

\subsubsection{Qatar-1}

Qatar-1\,b ($M_{\rm p}=1.29~M_{\rm J}$, $R_{\rm p}=1.14~R_{\rm J}$; \citeauthor{collins2017qatar1}~\citeyear{collins2017qatar1}) is the first planet discovered by the Qatar Exoplanet Transit Survey \citep{alsubai2011}. This HJ orbits a K3V dwarf star ($T_{\rm eff}=4861 \pm 25$~K) with $P_{\rm orb}=1.42$~d. In-transit and out-of-transit \jkt{RV}{radial velocity (RV)} measurements indicate a circular orbit within $2\sigma$ and a well-aligned configuration ($\lambda = -8.4^{\circ} \pm 7.1^{\circ}$); \citealt{covino2013}). Early timing analyses suggested possible orbital period variations \citep{vonessen2013}, but subsequent studies using extended baselines found no evidence for a changing period \citep{maciejewski2015,collins2017qatar1,mannaday2022}.

We found that $\Delta {\rm AIC} = 3.00$ and $\Delta {\rm BIC} = 6.54$, hence no convincing evidence of orbital decay from our linear and quadratic models. Our median quadratic model yields a quadratic coefficient 
$a = (0.00425^{+1.33}_{-1.28})\times10^{-11}\ \mathrm{d\ cycle^{-1}}$ which is consistent with zero ($|a|/\sigma_{\mathrm{avg}}\approx0.003$). The corresponding rate of period change is $0.0^{+0.58}_{-0.55}~ \mathrm{s\ yr^{-1}}$. Using the $3\sigma$ lower limit on the quadratic coefficient $a_1^{{\rm lim}} = -3.83 \times 10^{-11} ~{\rm d~cycle}^{-1}$, we therefore constrain the stellar tidal quality factor as $Q_\star^{\prime}> (1.16\pm0.09) \times 10^5$ in 99.7\% confidence level.

Theoretical models of Qatar-1 predict that gravity waves in the fully-damped regime would provide $Q^\prime_\mathrm{IGW}\approx 1.85\times 10^5$ at 11.6 Gyr. The planet exceeds the critical mass for wave breaking in this model ($0.5~M_{\rm J}$), which makes the likely operation of wave breaking a good justification for the fully-damped regime. We would thus predict the orbit to be decaying with $Q^\prime_\mathrm{IGW}\approx 1.85\times 10^5$. This is approximately at the same level as the lower bound observational constraint, making future observations particularly worthwhile.

We found two statistically significant peaks in the LS periodogram of the TTV diagram for Qatar-1 in the high-frequency regime (at 3.8 and 12.9~d, and two more in the low-frequency regime (at 50.5 and 613.8~d). However, our Keplerian fits based on these periods do not perform better than the linear fit. Qatar-1 is also known to be an active star. Although \cite{vonessen2013} reported a tentative periodic TTV signal at $\sim190$~d, we have not detected any peaks on our LS periodogram at the corresponding frequency. Later studies on the TTVs observed in Qatar-1 found different periodicities, although they are not statistically significant \citep{maciejewski2015,su2021,mannaday2022}. We believe that the frequencies that we and other studies have found through frequency analyses of the TTV data can be related to the activity of the star, which has been shown to be moderate in strength by \citet{covino2013} based on the emission profiles of Ca II H\&K lines in the HARPS data.

\subsubsection{Qatar-4}

Qatar-4\,b ($M_{\rm p} = 6.10 ~M_{\rm J}$ and $R_{\rm p}= 1.13 ~R_{\rm J}$; \citeauthor{alsubai2017}~ \citeyear{alsubai2017}) is an HJ orbiting around an early K-type star ($V=13.6$ mag and $T_{\rm eff}=5215$~K; \citeauthor{alsubai2017} ~\citeyear{alsubai2017}) in $P_{\rm orb} = 1.8$~d. The host-star is comparatively young with an age of $0.17$ ~Gyr \citep{alsubai2017}. The system has been subject to discussion for timing analyses only a few times \citep{mallonn2019,ivshinawinn2022} and no proof of a decaying orbit has been found yet.

Our analysis yielded $\Delta {\rm AIC} = -4.80$ and $\Delta {\rm BIC} = -2.74$. 
Hence, we conclude that there is no significant evidence for orbital decay with the current dataset. Our median orbital decay model suggests an orbital period change of $-59 \pm 7 ~{\rm ms~ yr}^{-1}$. Finally, we provide a constraint on the stellar tidal quality factor as $Q_\star^{\prime} > 1.1 \pm 0.4 \times 10^4$.

Our models of Qatar-4 indicate $Q^{'}_{\rm IGW}\approx 1.63\times 10^6$ at 0.15 Gyr if the gravity waves are fully damped, and that wave breaking is not predicted in the stellar core. This means that the fully-damped regime may not be appropriate, so it is unclear whether the planet's orbit should be decaying at the rate that would be predicted using this value, or whether tidal dissipation would be less efficient due to partial radiative damping of gravity waves. The star rotates too slowly for inertial waves to be excited by the planet.

In the LS periodogram of Qatar-4, there are two strong peaks at 29.6 and 47.3~d with FAP values smaller than $10^{-5}$. Qatar 4 displays out-of-transit variability due to its strong magnetic activity, as expected from its rather young age \citep{zak2025}. Based on the variability in WASP data, \citet{zak2025} found $7.07 \pm 0.08$~d for the rotation period. No non-zero eccentricity for its orbit, or evidence for an outside perturber, has been reported so far, making activity-induced pseudo-shifts in the transit centers the primary candidate for the reason of the peaks on the LS periodogram. 

\subsubsection{TOI-1937A}

TOI-1937A\,b is an ultra-short-period planet with $M_{\rm p}=2.01~M_{\rm J}$ and $R_{\rm p}=1.25~R_{\rm J}$, orbiting a Sun-like star ($T_{\rm eff}=5814$~K, $M_\star=1.072~M_\odot$, age $3.6^{+3.1}_{-2.3}$~Gyr) that is a member of a wide binary with a projected separation of 1030~au \citep{yee2023}.
\citet{jankowski2025} showed that the planet may undergo tidal engulfment within $\sim$500~Myr for $Q^{'}_\star < 10^7$ and within $\sim$1~Gyr for $Q^{'}_\star \approx 10^8$, while no engulfment is expected within 5~Gyr for $Q^{'}_\star \gtrsim 10^9$. Their TESS-based timing analysis revealed no statistically significant evidence for orbital decay or a perturber.

From our median models, we found $\Delta {\rm AIC} = 1.98$ and $\Delta {\rm BIC} = 5.05$, providing no significant evidence for a decaying orbit. Our median model for orbital decay claims a decreasing orbital period with a rate of $-5 \pm 11 {\rm ~ms ~yr}^{-1}$. Based on the measured orbital period change of TOI-1937A\,b, we place an upper limit on the stellar tidal quality factor of $Q^\prime_\star \lesssim 3.4 \times 10^9$ at the 95\% confidence level. The lower limit is constrained by the physical requirement $Q^{'}_\star > 0$ for tidal dissipation.

Our theoretical models of the star 
indicate that gravity waves would provide $Q^{'}_{\rm IGW}\approx 2.33\times 10^5$ at 3.6 Gyr, assuming the fully-damped regime. This is likely to be valid as the critical planetary mass for wave breaking is only $0.75 ~M_{\rm J}$, hence wave breaking is predicted and we would predict orbital decay to be occurring. This does not appear to be compatible with observations, so follow-up studies of this system would be worthwhile.

\subsubsection{TOI-2109}

TOI-2109\,b ($M_{\rm p} = 5.02~M_{\rm J}$, $R_{\rm p} = 1.347~R_{\rm J}$) has one of the shortest orbital periods among known exoplanets, orbiting in only 0.67~d ($\sim 16$~hrs), and is also among the hottest with $T_{\rm eq}=3631$~K \citep{wong2021}. This HJ orbits an F-type star with $M_\star=1.447~M_\odot$ and $T_{\rm eff}=6500$~K and has therefore been widely considered a prime target for studying orbital decay and tidally driven atmospheric escape \citep{wong2021,rosario2022,weinberg2024}. The likelihood of orbital decay was first evaluated by \citet{wong2021}, who predicted $\dot{P} \approx 10$--$740~ {\rm ms~ yr}^{-1}$ for $Q^{'}_\star \approx 10^5$--$10^7$. Using recent TESS and CHEOPS data, \citet{harre2024} reported tentative evidence for orbital decay after correcting the transit times with a sinusoidal TTV model, suggesting a nearby outer companion with $P_{\mathrm{c}} > 1.125$~days; however, confirmation is currently limited by the host star’s rapid rotation and the lack of precise RV constraints.

From our analysis, we found that $\Delta {\rm AIC} = -1.16$ and $\Delta {\rm BIC} = 1.09$. Hence, we conclude that no statistically supported evidence for orbital decay has been found. From our quadratic model, we measured an orbital change of $26 \pm 11 ~\text{ms yr}^{-1}$. Again, using the $3\sigma$ lower limit of $dP/dE$, we constrain the modified stellar tidal quality factor $Q_\star^{\prime} > (1.23 \pm 0.20) \times 10^6$. 

The star probably rotates rapidly with an estimated rotation period of approximately $P_\mathrm{rot}=2\pi R_\star(v \sin{i})^{-1} = 1.14$~d adopting $v\sin{i}= 81.2 ~\mathrm{km~s^{-1}}$ \citep{wong2021}. This is sufficiently fast that inertial waves may be excited in the stellar convective envelope. Our models for the star predict $Q^{'}_\mathrm{IGW}\approx 1.2\times 10^8$ for gravity waves (assuming the fully-damped regime, which is not expected to be valid as wave breaking is not predicted) at 1.77 Gyr, and $Q^{'}_\mathrm{IW}\approx 6.9\times 10^6$ at the same age due to inertial waves -- which is compatible with the observational constraints. Thus, inertial waves are probably the dominant tidal mechanism in this system, and since $P_\mathrm{orb}<P_\mathrm{rot}$, we would predict orbital decay (and stellar spin-up) driven by these waves. Note that the value adopted here gives a representative value for inertial wave dissipation across the whole frequency range in which these waves can be excited, and the actual value we would expect for the tidal frequency of this system is more uncertain. (It could be larger or smaller, depending on the proximity of the tidal frequency to the frequency of the most dissipative inertial mode peaks in the response \citep[e.g.][]{ogilvie2013,astoul2023}.)

\subsubsection{TrES-1}

TrES-1\,b ($M_{\rm p}= 0.752~M_{\rm J}$, $R_{\rm p} = 1.067~R_{\rm J}$; \citeauthor{torres2008}~\citeyear{torres2008}) is one of the first exoplanets discovered by the transit method \citep{alonso2004} and therefore has the longest baseline in our orbital decay sample, with more than 20 years of data. It orbits a K0 main-sequence star with $T_{\rm eff}=5230$~K and $M_\star=0.878~M_\odot$ \citep{bonomo2017} in $\sim$3~d. Although it was not initially included due to its different location in parameter space (Fig.~\ref{fig:tau_comparison}), we manually involved this system given the extensive attention it has received in the literature. Early timing analyses found no significant TTVs \citep{rabus2009,baluev2015}. After the inclusion of TESS data, \cite{ivshinawinn2022} reported a period decrease of $-18.36 \pm 3.73~{\rm ms~ yr}^{-1}$. Using extended ETD coverage, \cite{hagey2022} found support for orbital decay with $\Delta {\rm BIC}=-9.7$ and $dP/dt=-10.9 \pm 2.1~{\rm ms~ yr}^{-1}$. More recently, \cite{adams2024} derived a decay rate of $-16.4 \pm 4.9~{\rm ms~yr}^{-1}$ over 2293 cycles ($\Delta {\rm BIC}=-68.8$), but inferred a physically implausible tidal quality factor of $Q^{'}_\star=160$.

From our timing analysis, we also found that the orbit is likely to be undergoing decay as the goodness of fit favors the quadratic model with $\Delta {\rm AIC}=-13.57$ and $\Delta {\rm BIC}=-11.02$. Our median quadratic model indicates a decrease in the orbital period of TrES-1\,b with a rate of $-14.9\pm0.6 {\rm ~ms~ yr}^{-1}$, in agreement with the previous work within the 95\% confidence level. Similar to \cite{adams2024}, we infer a value of $Q^{'}_\star = 570 \pm 60$, which is about three orders of magnitude lower (i.e., more efficient) than that of WASP-12\,b and at least five orders of magnitude lower than the theoretical predictions of \cite{weinberg2024}, who estimated $Q_\star^{\prime} > 10^7$. For comparison, WASP-12 is expected to be a subgiant star and exhibits efficient tidal dissipation via IGW (see \ref{subsub:wasp12}). However, as indicated by \cite{adams2024} too, there are no indications for TrES-1 to be in its subgiant phase. Therefore, although the observed timing variations are statistically significant, the inferred $Q_\star^{'}$ is not in agreement with standard tidal dissipation models, and alternative mechanisms may be responsible for the observed signal.

Our theoretical models of TrES-1 
indicate that gravity waves provide $Q^{'}_\mathrm{IGW}\approx 2.57\times 10^6$ at 3.7 Gyr. The critical planetary mass is $15.4~M_{\rm J}$, suggesting that wave breaking is not predicted in the radiative core of the star. The star rotates too slowly for inertial waves to be excited for an aligned orbit and equilibrium tide dissipation is also predicted to be negligible. Hence, we would predict the orbit to be decaying slowly, with $Q^{'}_\mathrm{IGW}\gtrsim 2.57\times 10^6$, and possibly much more slowly than this unless the system is lucky enough to resonantly excite a g-mode. If this is the case, it is possible in principle for $Q^{'}$ to be smaller, though it is likely to be difficult to reconcile with the observational constraint.

There are two statistically significant peaks on the LS periodogram of the TTV data for TrES-1\,b. The first one corresponds to 37.4~d and the second to 54.2~d with FAP values smaller than 0.01\%. The potential companion at 13.16 arcseconds \citep{michel2024} cannot be the source of such a signal because it is too far away, although it is most probably gravitationally bound. 
A potential TrES-1\,c has been suggested to cause an increase in the flux during a transit of TrES-1\,b by \citet{rabus2008}, and in a long-term variation in the radial velocities by \citet{hagey2025}, most recently, which cannot induce the observed TTVs due to its long period ($\sim1200$~d). However, neither claims were not supported by observational evidence as definite causes. Observations of flux variations due to magnetic activity have been reported, which might have caused a flux increase if brighter regions (faculae) were involved. Such magnetic activity-induced variations are also observed to affect RV observations. \citet{hagey2025} investigated the potential of apsidal motion, but the amplitude of their TTV diagram can only be explained by the presence of an undetected close-in planetary companion because the orbit of TrES-1\,b was found to be circular. Therefore, we think that both claims of potential companions and the statistically significant frequencies in our LS periodogram might be related to the same source, the magnetic activity of the star.

\subsubsection{WASP-3}

WASP-3\,b ($M_{\rm p} = 1.89~M_{\rm J}$, $R_{\rm p} = 1.42~R_{\rm J}$; \citeauthor{bonomo2017}~\citeyear{bonomo2017}) is a HJ orbiting a F7--8 type ($T_{\rm eff}=6400$~K) main-sequence star ($2.1 \pm 1.2$~Gyr; \citeauthor{southworth2011}~\citeyear{southworth2011}) in 1.8~d, discovered by \cite{pollacco2008}, with a slightly misaligned orbit ($\lambda = 15^{\circ} \pm 10^{\circ}$; \citeauthor{simpson2010}~\citeyear{simpson2010}). The first discussion of the tidal evolution of the system was presented by \cite{pont2009}. The system has been extensively monitored for orbital period variations. While some studies reported deviations from a constant period \citep{maciejewski2010wasp3,eibe2012,nascimbeni2013wasp3}, others found no evidence for inner or outer perturbers \citep{montalto2012,maciejewski2013wasp3,maciejewski2018wasp3}.

As a result of our TTV analysis, we found $\Delta \text{AIC}=1.32$ and $\Delta \text{BIC}=4.64$. Thus, we conclude that there is no evidence to indicate that the system is undergoing orbital decay. From our median quadratic model, we found $dP/dt = -1.6\pm0.8 ~\text{ms~yr}^{-1}$ and this corresponds to $Q_\star^{\prime} > (3.8 \pm1.3) \times 10^6$ at 95\% confidence level.

Our theoretical models indicate that efficient gravity wave dissipation could provide 
$Q^{'}_\mathrm{IGW}\approx 3.8\times 10^7$ at 2.1 Gyr, in the fully-damped regime. The star has a convective core, so wave breaking is not predicted, making it uncertain whether the fully-damped regime is applicable \jkt{\citep[though magnetic wave conversion is another possibility to justify it,][]{duguid2024}.}{(though magnetic wave conversion is another possibility).}


\subsubsection{WASP-12}
\label{subsub:wasp12}

WASP-12\,b might be the only exception in our sample, as our primary goal was not to detect orbital decay itself, but to update relevant parameters and discuss the host star's tidal dissipative efficiency. First discovered by \citet{hebb2009}, WASP-12\,b ($M_{\rm p} = 1.47~M_{\rm J}$, $R_{\rm p} = 1.96~R_{\rm J}$; \citeauthor{collins2017qatar1}~\citeyear{collins2017qatar1}) was the first exoplanet with observed tidal orbital decay\jkt{ for over a decade}{}. This HJ orbits a F9 star ($T_{\rm eff} = 6360_{-140}^{+130}$~K; \citeauthor{collins2017qatar1}~\citeyear{collins2017qatar1}) in 1.09~d. \citet{maciejewski2016} reported a period decrease of $dP_{\rm orb}/dE = (-8.9 \pm 1.4) \times 10^{-10}~{\rm d~ cycle}^{-1}$, confirmed by subsequent studies \citep{patra2017,maciejewski2018,ozturkerdem2019,yee2020,turner2021,wong2022,kutluay2023,alvarado2024,adams2024,leonardi2024,sodickson2025}. Debate focuses on the host star's evolutionary state: \cite{weinberg2017} and \cite{barker2020} suggest a subgiant to explain $Q^{'}_\star \sim 10^5$ via wave breaking, whereas \cite{duguid2024} argue magnetic wave conversion in a main-sequence star with a convective core could also explain the observed $Q^{'}_\star$. Overall, our goal was to provide updated ephemerides and $Q_\star^{\prime}$.

Building upon the previous arguments, our timing analysis also supports a decaying orbit for WASP-12\,b, with $\Delta \mathrm{AIC} = -189.7$ and $\Delta \mathrm{BIC} = -186.25$. From our median quadratic fit, we calculated an orbital period decrease of $-29.4 \pm 4.0~\mathrm{ms~yr^{-1}}$, which agrees with previous estimates within the 95\% confidence level. Indeed, the most recent calculations by \cite{adams2024} and \cite{alvarado2024}, found $dP/dt= -29.8 \pm 1.6~\mathrm{ms~yr^{-1}}$ and  $dP/dt = -29 \pm 2~\mathrm{ms~yr^{-1}}$, respectively. The corresponding modified stellar tidal quality factor is $Q_\star^{\prime} = (1.72 \pm 0.18) \times 10^5$.
In addition, we also fitted a cubic model to provide an estimate of the rate of tidal decay of WASP-12\,b. Following \cite{alvarado2024}, the acceleration of the orbital period change in this model can be calculated from the cubic model, defined as
\begin{equation}
    T(E) = T_0 + P_{\rm orb}\times E + \frac{1}{2}\frac{dP_{\rm orb}}{dE} \times E^2 + \frac{1}{6}\frac{d^2P_{\rm orb}}{dE^2}\times E^3,
\label{eq:cubic}
\end{equation}
and the corresponding rate of change of the rate of tidal decay is analytically provided as 
\begin{align}
\ddot{P} &\approx 1.2 \times 10^{-24}~{\rm s^{-1}}
\left( \frac{Q^\prime_\star}{10^6} \right)^{-1}
\left( \frac{M_\mathrm{p}}{M_\mathrm{J}} \right)
\left( \frac{M_\star}{M_\odot} \right)^{-8/3} 
\nonumber\\
&\quad \times
\left( \frac{R_\star}{R_\odot} \right)^5
\left( \frac{P}{\rm d} \right)^{-13/3}
\left( \frac{\dot{P}}{10^{-9}} \right).
\end{align}

\begin{figure}
    \centering
    \includegraphics[width=0.39\textwidth]{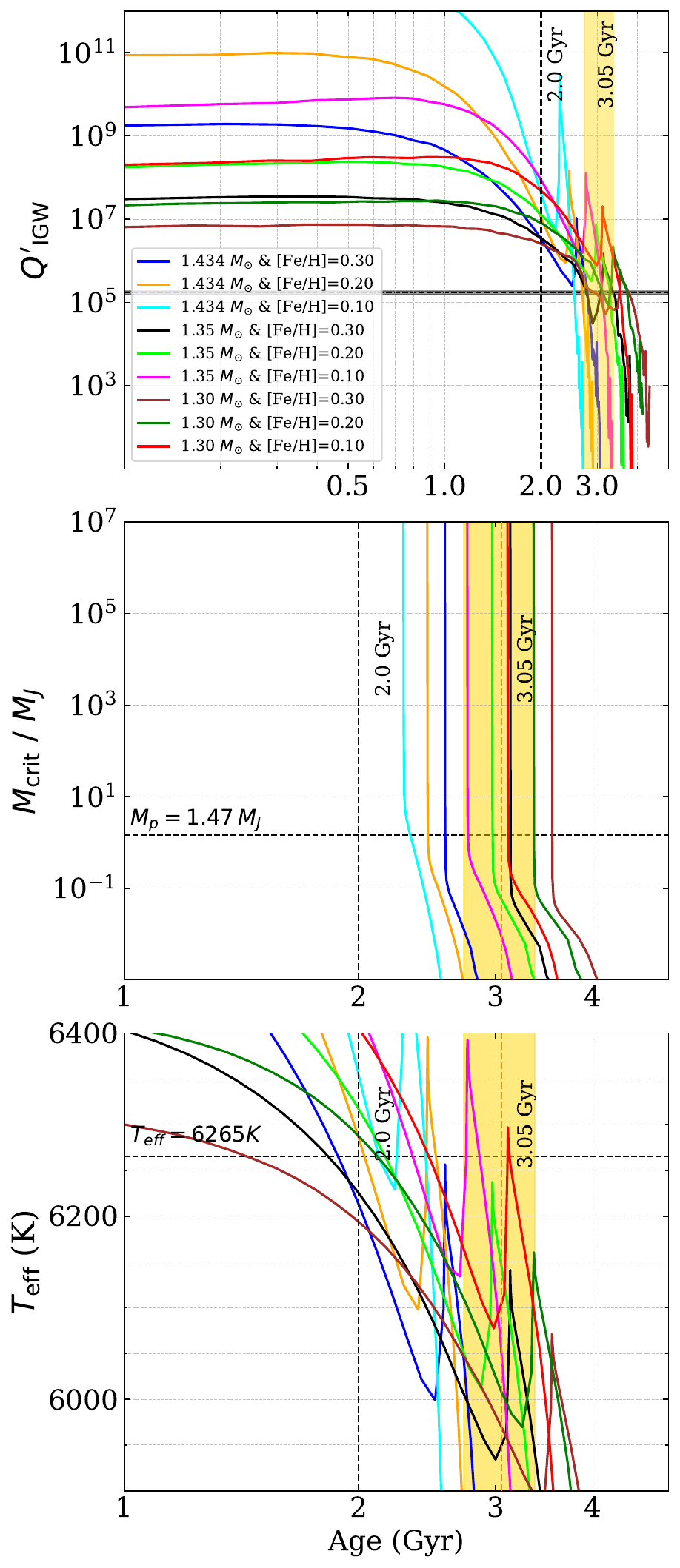}
    \caption{Top: WASP-12's tidal efficiency due to IGW $(Q_\text{IGW}^{'})$ as a function of stellar age for 9 different stellar evolution models. The dashed horizontal line displays $Q_\star^{\prime} = 1.72 \times 10^5$ from our median quadratic model. Middle: Critical mass of WASP-12\,b for wave breaking in the stellar core $M_\text{crit}$ as a function of stellar age. The horizontal dashed line shows the planetary mass $M_{\rm p} = 1.47 ~M_{\rm J}$ from \citet{collins2017qatar1}. Bottom: WASP-12's stellar effective temperature $T_\text{eff}$ as a function of stellar age. Here, dashed horizontal line shows $T_\text{eff} = 6265 ~K$ from \citet{leonardi2024}. For each plot, we show the age reported by \citet{collins2017qatar1} (2.0 Gyr) as a vertical black dashed line, and the age from \citet{leonardi2024} ($3.05 \pm 0.22$ Gyr) as an orange dashed line at the nominal value. The gold-shaded region represents the 68\% confidence interval.} 
    \label{fig:wasp12_models}
\end{figure}

\begin{figure*}
    \centering
    \includegraphics[width=\textwidth]{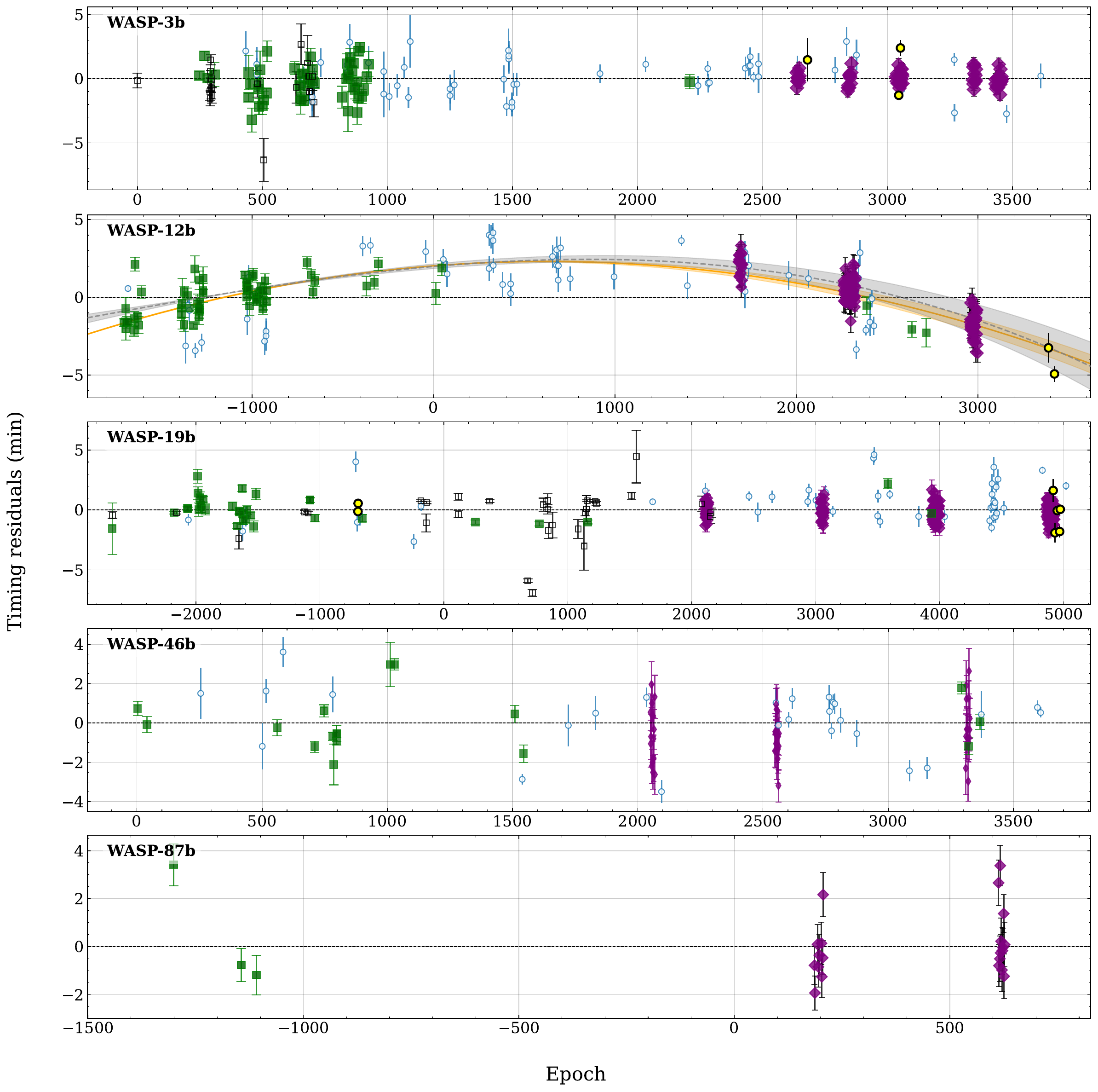}
    \caption{Same as Figure \ref{fig:ttv_plot_1}, but for WASP-3\,b, WASP-12\,b, WASP-19\,b, WASP-46\,b and WASP-87\,b. Since $|\Delta \text{BIC}| \gg 10$ for WASP-12\,b case, we present the median orbital decay model represented with the orange line and the shaded band indicating its $3\sigma$ uncertainty range. Additionally, we show the median of the decay acceleration model in black dashed lines and its $3\sigma$ uncertainty range in a grey band.}
    \label{fig:ttv_plot_3}
\end{figure*}

\noindent Again, we followed the same methodology described in \S~\ref{sec:ttv_analysis} while implementing the probabilistic fitting. Our cubic fit is displayed in Fig.~\ref{fig:ttv_plot_3}. From our median acceleration model, we found that  $d^2P_\text{orb}/dE^2 = (-3.81\pm 0.52) \times 10^{-13} ~\text{d cycle}^{-2}$. Hence, the corresponding acceleration of orbital decay in WASP-12\,b is $\ddot{P} \approx -4.29 \times 10^{-23} ~\text{s}^{-1}$, which is slightly faster than what \cite{alvarado2024} found earlier. This difference might be considered quite negligible when the change is also negligibly small, so we consider our results to be consistent with those of \citet{alvarado2024}. 

From our theoretical models (Fig. \ref{fig:wasp12_models}), we find that IGWs can provide $Q_\star^{\prime} \approx 10^5$ for an age of approximately 3.0~Gyr for all models with stellar masses of $\{1.30, 1.35, 1.434\}\,M_\odot$ and metallicities of $\{0.1, 0.2, 0.3\}$ dex. The predicted critical planetary mass at this age appears to be lower than the observed value of $M_{\rm p} = 1.47\,M_{\rm J}$~\citep{collins2017qatar1}. Our models, therefore, suggest that the host star should currently be in its subgiant phase in order to reproduce the observed value 
$Q_\star' = (1.72 \pm 0.18) \times 10^5$, consistent with the findings of 
\citet{barker2020}. Alternatively, \citet{duguid2024} showed that the same fully-damped regime for gravity waves can be explained by magnetic wave conversion even if the star is a main-sequence star with a convective core, which remains another possibility for this system.
On the other hand, IWs are not expected to be excited (by the asynchronous tide with $l=m=2$) in the convective zone because the stellar rotation period is $P_{\text{rot}} = 37.4~\text{days}$, as inferred from $v\sin i = 2.2~\text{km\,s}^{-1}$~\citep{bonomo2017} and does not satisfy the criteria mentioned in \S~\ref{sec:introduction}.

\subsubsection{WASP-19}

WASP-19\,b ($M_{\rm p} = 1.154~M_{\rm J}$, $R_p = 1.415~R_{\rm J}$; \citeauthor{corteszuleta2020}~\citeyear{corteszuleta2020}) was the shortest-period exoplanet when it was discovered \citep{hebb2010wasp19} and is the second shortest in our sample after TOI-2109\,b, making it a strong candidate for tidally-driven orbital decay. It orbits a G8 star ($T_\text{eff}=5616$~K, $M_\star=0.965~M_\odot$, $R_\star=1.006~R_\odot$) in 0.78~d, with a well aligned orbit ($\lambda=-4\overset{\circ}{.}6 \pm 5\overset{\circ}{.}2$; \citeauthor{hellier2011}~\citeyear{hellier2011}). Since discovery, the orbital decay scenario has been probed extensively (e.g., \citealt{mancini2013}). \citet{petrucci2020} found a linear ephemeris favored the quadratic model with $Q_\star^{\prime} > (1.23 \pm 0.23) \times 10^6$. Reported period changes vary: $-6.5 \pm 1.3$ (\citeauthor{patra2020}~\citeyear{patra2020}) and $-3.7 \pm 0.5~\text{ms yr}^{-1}$ (\citeauthor{korth2023}~\citeyear{korth2023}). \citet{maciejewski2024} found no significant signal, but \citet{sodickson2025} reported $-3.89 \pm 0.37~\text{ms yr}^{-1}$. Recent studies favor apsidal motion ($\Delta \text{BIC}=-370.4$; \citeauthor{biswas2024}~\citeyear{biswas2024}) and we will discuss this scenario in \S~\ref{subsec:apsidalmotion}.

Our median models yielded $\Delta \mathrm{AIC} = 1.96$ and $\Delta \mathrm{BIC} = 5.63$, providing no evidence of decreasing or increasing orbital period over time. Using the 1.96$\sigma$ lower limit on the quadratic coefficient is $a_\text{lower} = -6.884 \times 10^{-12} \text{~d cycle}^{-1}$, we provide $Q_\star^{\prime} > 5.0 \pm 0.7 \times 10^6$ at 95\% confidence level. 

Our theoretical models of WASP-19 predict that fully-damped gravity waves provide $Q^\prime_\mathrm{IGW}\approx 6 \times 10^4$ for ages of approximately 6 Gyr (ranging from $8\times 10^4$ to $5\times 10^4$ for ages between 3 and 7 Gyr, respectively). 
The critical planetary mass for wave breaking passes below $1~M_{\rm J}$ for stellar ages of approximately 6 Gyr. This indicates that the fully damped regime is very likely for ages older than this, as this value exceeds the planetary mass. It is therefore difficult to explain the inferred $Q^\prime_\star>5\times 10^6$ unless the stellar age is less than 6 Gyr. The critical mass is as large as $10~M_{\rm J}$ for 3 Gyr ages, decreasing to $1~M_{\rm J}$ by 6 Gyr. We can therefore plausibly explain the observational constraint if the star is younger than 6 Gyr. Inertial waves cannot be excited by an aligned orbit, and equilibrium tide damping is likely to be negligible. This theoretical interpretation is very similar to the one presented in \citet{barker2020}. 

WASP-19\,b's orbit was found to be slightly eccentric (e = $0.0126^{+0.0140}_{-0.0089}$; \citeauthor{corteszuleta2020}~\citeyear{corteszuleta2020}). \citet{biswas2024} proposed that its apsidal motion is causing TTVs, which might be amplified by magnetic activity. Recent studies have not found any outside perturber \citep{petrucci2020,sodickson2025} that would cause the observed TTVs. We found a significant periodicity at 16.75~d, which can only be associated with the magnetic activity of the star. Some of the transit light curves of WASP-19\,b are heavily affected by starspot-induced asymmetries. Out-of-transit variability is evident in the TESS light curves as well. When the data are pre-whitened from this peak, another peak is observed at lower frequencies, corresponding to $\sim1536$~d. However, our sinusoidal fits with this periodicity do not compete with the linear model in terms of fit statistics. When we phase-fold the data with the periodicity, we do not observe a good agreement with the sinusoidal model either. Nonetheless, we report these frequencies for future work on the TTVs of the target. 

\subsubsection{WASP-46}

WASP-46\,b ($M_{\rm p} = 1.91~M_{\rm J}$, $R_{\rm p} = 1.174~R_{\rm J}$; \citeauthor{ciceri2016}~\citeyear{ciceri2016}) is an HJ orbiting a G6 star ($T_\text{eff} = 5600$~K) in a 1.43-day circular orbit, discovered by \citet{anderson2012wasp46}. The host star shows chromospheric activity, allowing age estimates of 1.5~Gyr \citep{anderson2012wasp46} and later 9.6~Gyr \citep{ciceri2016,bonomo2017}. For the investigation of orbital decay, \citet{petrucci2018wasp46} used 6 years of photometry and found that orbital decay model favors ($\Delta \text{BIC} = -5.0$) with a stellar tidal quality factor $Q_\star^{\prime} > 7 \times 10^3$ from $\delta P = -0.12 \pm 0.45~\text{ms yr}^{-1}$. \citet{adams2024}, using nearly 14 years of data, found a period increase of $21.6 \pm 8.2~\text{ms yr}^{-1}$, but after rescaling errors and removing outliers, $\Delta\text{BIC} = 2.0$. \jkt{, highlighting the need for further observations due to murky early light-curve data.}{This highlights the need for further observations because the early light-curve data are not of high quality.}

Our timing analysis resulted in $\Delta \mathrm{AIC} = -1.13$ and $\Delta \mathrm{BIC} = 1.36$. Therefore, in agreement with \citet{petrucci2018wasp46} and \citet{adams2024}, we also found no significant evidence for orbital decay signal in TTV data. From the median orbital decay model, the 3$\sigma$ period derivative values are 1.71$\times10^{-10} \text{~d cycle}^{-1}$ and 3.10$\times10^{-10} \text{~d cycle}^{-1}$. Since both values are positive, the corresponding tidal quality factors would be negative. Because $Q^\prime_\star$ must be physically positive (for tidal dissipation rather than anti-dissipation), no meaningful 3$\sigma$ lower or upper limit can be derived from our analysis. 

Our theoretical models for the host star 
find $Q^{'}_\mathrm{IGW}\approx 2.58\times 10^5$ for gravity waves in the fully-damped regime at 9.6 Gyr. We estimate that the critical planetary mass for wave breaking is $1.7~M_{\rm J}$ at the same age, which is slightly lower than the mass of the planet, indicating that wave breaking is predicted -- though this is marginal. We would predict orbital decay at a rate consistent with $Q{'}_\mathrm{IGW}\gtrsim 2.58\times 10^5$, with the lower limit achievable if the waves are fully damped.

\subsubsection{WASP-87}

WASP-87\,b is a HJ orbiting a metal-poor ([Fe/H] = $-0.41\pm0.10$) mid-F star ($T_\text{eff} = 6250\pm110$ K) with $P_\text{orb} = 1.68$~d \citep{anderson2014}. Rossiter–McLaughlin observations indicate a nearly aligned orbit ($\lambda = -8^{\circ}\pm11^{\circ}$; \citealt{addison2016}), and the host star has a rotation period of $\sim8.6$~d. The system has a Sun-like companion, WASP-87\,B, with both stars having similar ages ($3.8 \pm 0.8$ and $ 3.8 \pm 0.6$ Gyr; \citealt{anderson2014}). \emph{Spitzer} occultation observations confirm a circular orbit \citep{2020AJ....159..137G}. WASP-87 lies in the sparsely populated $P_\text{rot}$–$P_\text{orb}$ region \citep{2013ApJ...775L..11M}, potentially shaped by tidal interactions.

As a result, the 3$\sigma$ period derivative values are $2.34\times10^{-10}$ and $2.14 \times 10^{-9}$, both in d cycle$^{-1}$ unit. Since both values are positive, the corresponding tidal quality factors would be negative. Because $Q^\prime_\star$ must be positive, no meaningful 3$\sigma$ lower or upper limit can be derived from these data. We detect no significant TTVs but, by updating the ephemeris, we note that the bright, hot host star makes the system a good target for atmospheric studies at shorter wavelengths \citep{2017AJ....153...81C}.

Our theoretical models predict $Q{'}_\mathrm{IGW}\approx4.5\times 10^6$ at 1 Gyr, due to gravity wave dissipation if these waves are fully damped. However, wave breaking is not predicted in the star, probably making this a lower bound on $Q^{'}_\star$.

\subsubsection{WASP-103}

WASP-103 b is one of the hot Jupiters in our sample with an orbital period shorter than one day, and it has received considerable attention since its discovery by \citet{gillon2014}. Although the system lies near the ecliptic and therefore lacks TESS coverage, ground-based efforts have provided a nearly uninterrupted and long-term observational baseline up to the present day. \citet{basturk2022} first investigated the possibility of a changing orbital period but found no evidence. Similarly, \citet{maciejewski2022} found $\Delta {\rm BIC} = 3.0$ and \citet{barros2022} stated that the linear ephemeris is preferable for the current set of light curves. \jkt{Most}{The most} recent attempt came from \citet{alvarado2024} but again no compelling evidence supporting a period variation was reported.

As in previous studies, our analysis also fails to identify a statistically preferable model for a changing orbital period. From our median quadratic model, we find $\dot{P} = -2.0 \pm 0.7~{\rm ms~yr^{-1}}$ and we derive a lower limit of $Q_*' > (1.83 \pm 0.17) \times 10^6$ at the 95\% confidence level.

Our theoretical models predict $Q^{'}_\mathrm{IGW}\approx 1.8\times 10^6$ at 1.7 Gyr due to gravity waves, which is remarkably close to the observational lower limit. Wave breaking is not predicted though, so this is probably a lower bound on the dissipation due to gravity waves, unless the system happens to excite a g-mode in resonance.

\subsubsection{WASP-114}

WASP-114\,b is a neglected planet in terms of its potential to display orbital decay, although it is a massive (1.769 M$_{\rm J}$) planet on a very short-period  (P$_{\rm orb}$ = 1.5488~d) orbit \citep{barros2016}. \citet{patra2020} constructed the first ever TTV diagram with only two data points spanning less than a thousand epochs. Later \citet{kokori2022}, \citet{ivshinawinn2022}, and \citet{kokori2023} only updated its linear ephemerides.

From our median models, we found $\Delta \mathrm{AIC} = -0.56$ and $\Delta \mathrm{BIC} = 0.94$. Therefore, we concluded that from the TTV analysis, there is no statistically significant evidence for a changing orbital period. From our median quadratic model, we found a decreasing orbit with a rate of $-35 \pm 10 \rm ~ms~yr^{-1}$. So, we used this value to constrain stellar tidal quality factor and we provide $Q_*^{\prime} > 4.7 \pm 0.6 \times 10^6$ at the 95\% confidence level.

Our theoretical models 
predict $Q^{'}_\mathrm{IGW}\approx 4.8\times 10^3$ at 4.3 Gyr due to gravity waves. Since the star is likely to be a post-MS star, perhaps in the sub-giant phase, the star has a radiative core and wave breaking is predicted for the planetary mass, as $M_{\rm crit} < 0.1~M_{\rm J}$ by the estimated age of the star. Hence, we predict orbital decay for the planet at a rate consistent with $Q^{'}_\mathrm{IGW}\approx 4.8\times 10^3$, though the precise value is sensitive to stellar age during this phase of the star's evolution. For the models we have considered, the theoretical predictions appear to be in conflict with the observational constraint, though the predictions do depend strongly on stellar age.

\begin{table*}
\centering
\caption{Median-fit results from our TTV analysis, presented for both the linear model representing the constant period and the quadratic model that accounts for orbital decay. In cases where $dP/dE$ (or equivalently $dP/dt$) is positive---indicating orbital growth---the \(3\sigma\) lower limit is adopted to derive a lower bound on reduced tidal quality factor $Q_\star^{\prime}$ of the host star at 99.7\% confidence level. It should be noted that the uncertainties on $Q_\star^{\prime}$ come from the uncertainties on other parameters given in Eq. \ref{eq:modified_tidal_quality_factor}. }
\label{tab:ttv_results}
\renewcommand{\arraystretch}{1.3} 
\begin{tabular}{lccccccc}
\hline
& & & & & \multicolumn{3}{c}{\textbf{Quadratic Model}} \\
\cline{6-8}
System & $\Delta \mathrm{AIC}$ & $\Delta \mathrm{BIC}$ & $P_{\mathrm{orb}}$\textsuperscript{\hyperref[fn:c]{c}} & $T_0$\textsuperscript{\hyperref[fn:b]{b,}}\textsuperscript{\hyperref[fn:c]{c}} & $a_\text{quad}$\textsuperscript{\hyperref[fn:d]{d}}& $d{P_\text{orb}}/dt$ &$Q_\star'$\textsuperscript{\hyperref[fn:a]{a}} \\
&  &  & (days) &  (BJD$_{\rm TDB}$) & (d cycle$^{-1}$) & (ms yr$^{-1}$) & \\
\hline
CoRoT-2 & 1.93 & 4.97 & 1.7429971349(2) & 57683.441534(37) & $3.14^{+1.61}_{-1.69} \times 10^{-11}$ & $1.14 \pm 0.58$ & $> 5.27 \pm 0.87 \times 10^5$ \\
HAT-P-23 & 1.77 & 4.60 & 1.2128864160(2) & 57742.573862(29) & $1.53^{+1.33}_{-1.32} \times 10^{-11}$ & $0.80 \pm 0.69$ & $> 1.41 \pm 0.45 \times 10^5$ \\
HATS-18 & 1.72 & 4.65 & 0.8378438220(45) & 58626.511345(37) & $3.06^{+3.29}_{-3.31} \times 10^{-11}$ & $2.31\pm2.48$ & $> 6.74 \pm 2.04\times10^5$ \\
KELT-9  & 1.97 & 4.75 & 1.4811189635(34) & 57095.685686(62) & $1.06^{+2.25}_{-2.60} \times 10^{-11}$ & $0.45\pm0.96$ & $> 1.21\pm 0.37 \times 10^6$ \\
KELT-16 & $-0.68$ & 1.88 & 0.9689926792(51) & 58392.597911(42) & $-2.57_{-0.68}^{+0.68} \times 10^{-10}$ & $-16.76 \pm 4.48$ & $> 1.03\pm 0.21 \times 10^6$\\
Qatar-1 & 3.00 & 6.54 & 1.4200242456(12) & 56234.103712(21) & $< 10^{-13}$ & $<0.006$ & $> 1.16 \pm 0.09 \times 10^5$ \\
Qatar-4        & $-4.80$ & $-2.74$ & 1.8053646151(39) & 57637.774877(13) & $-1.68^{+0.21}_{-0.20} \times 10^{-9}$  & $-59 \pm 7$ & $> 1.10 \pm 0.40 \times 10^4$ \\
TOI-1937A  & 1.98 & 5.05 & 0.9466794627(95) & 59085.910210(96) & $-0.73^{+1.71}_{-1.67} \times 10^{-10}$ & $-5 \pm 11$ & $< 3.4 \pm 0.1 \times 10^9$\\
TOI-2109 & -1.16 & 1.09 & 0.6724740164(57) & 59378.459298(46) & $2.77_{-1.12}^{+1.13} \times 10^{-10}$ & $26 \pm 11$ & $> 1.23 \pm 0.20 \times 10^6$ \\
TrES-1 & $-13.57$ & $-11.02$ & 3.0300696522(27) & 56822.891950(88) & $-7.14_{-2.86}^{+2.86}\times10^{-11}$ & $-14.9\pm0.6$ & $5.7 \pm 0.6 \times 10^2$ \\
WASP-3  & 1.32 & 4.64 & 1.8468351793(20) & 54143.851121(47) & $-4.82^{+1.89}_{-2.26} \times 10^{-11}$ & $-1.6 \pm 0.8$ & $> 3.8 \pm 1.3 \times 10^6$ \\
WASP-12 & $-189.70$ & $-186.25$ & 1.0914193645(14) & 57010.513047(20) & $-5.08^{+0.01}_{-0.01} \times 10^{-10}$ & $-29.4 \pm 4.0$ & $1.72 \pm 0.18 \times 10^5$ \\
WASP-19 & 1.96 & 5.63 & 0.7888390012(58) & 56885.482609(15) & $-2.27^{+2.36}_{-2.44} \times 10^{-12}$ & $-0.18 \pm 0.19$ & $> 5.0\pm0.7 \times 10^6$ \\
WASP-46 & -1.13 & 1.36 & 1.4303721864(35) & 55392.316213(75) & $2.36^{+0.24}_{-0.22} \times 10^{-10}$ & $10.4 \pm 1.1$ & -- \\
WASP-87 & $-1.51$ & $-0.33$ & 1.6827943515(95) & 58276.860909(122) & $1.20^{+0.21}_{-0.21}\times 10^{-9}$ & $47 \pm 8$ & -- \\
WASP-103 & 0.68 & 3.12 & 0.9255454474(24) & 57308.324556(16) & $-5.30_{-2.01}^{+2.03} \times 10^{-11}$ & $-2.0 \pm 0.7$ & $>1.83 \pm 0.17 \times 10^6$ \\
WASP-114 & $-0.56$ & $0.94$ & $1.5487751950(12)$ & 57522.659955(123) & $-5.10^{+1.53}_{-1.53} \times 10^{-10}$ & $-35 \pm 10$ & $>4.7 \pm 0.6\times10^5$ \\
WASP-121 & $-16.82$ & $-13.84$ & 1.2749244026(31) & 58119.720671(16) & $3.70_{-0.22}^{+0.21} \times 10^{-10}$ & $15.1 \pm 0.8$ & -- \\
WASP-122 & $-1.13$ & $1.32$ & $1.7100531597(69)$ & 56665.225124(123) & $-1.65^{+0.59}_{-0.63} \times 10^{-10}$ & $-19\pm7$ & $>4.6\pm0.5 \times10^4$ \\
WASP-167 & {$-3.81$} & {$-5.4$} & 2.0219566280(13) & 58117.023535(109) & {$3.8^{+1.3}_{-1.3} \times 10^{-10}$} & {$11.8 \pm 4.1$} & {$> 10.9 \pm 0.11 \times 10^7$} \\
\hline
\hline
\multicolumn{8}{l}{
\parbox{\linewidth}{\footnotesize{ 
\label{fn:a}$^\text{a}$ The 3$\sigma$ lower limit on $Q^{'}_\star$ is calculated from the lower limit of $\dot{P}$ (or $dP/dE$); no value is provided if $\dot{P}-3\sigma$ is positive. \par
\label{fn:b}$^\text{b}$ $\rm BJD_{TDB}$ - 2,400,000. \par
\label{fn:c}$^\text{c}$ We provide $T_0$ and $P_{\rm orb}$ according to the quadratic ephemerides for TrES-1\,b, WASP-12\,b and WASP-121\,b since their quadratic ephemeris represents the timing data better. \par
{{\label{fn:d}$^{\text{d}}$ }$a_{\mathrm{quad}} = \tfrac{1}{2}\,\frac{dP}{dE}$.}}}}
\end{tabular}
\end{table*}

\subsubsection{WASP-121}

WASP-121\,b attracted attention because it exhibited a radius anomaly (1.865 $R_{\rm J}$ for 1.183 $M_{\rm J}$) and also due to its proximity to the Roche limit, which is only 1.15 times its current orbital separation \citep{delrez2016}. \citet{salz2019} reported tentative ($1.9\sigma$) ultraviolet absorption of 0.55\%, potentially due to atmospheric mass loss, similar to WASP-12\,b \citep{fossati2010}, and \citet{yan2021} derived a mass-loss rate of $1.28 \times 10^{12}$ g s$^{-1}$. The planet has an almost polar orbit \citep{bourrier2020}, which is possibly linked to its late arrival and minimal damping of orbital obliquity \citep{spalding2022}. \citet{maciejewski2022a} compiled mid-transit times from literature and TESS Sectors 7, 33, and 34, showing no departure from linear ephemerides. \citet{adams2024} analyzed 66 data points and noted that only the earliest composite point suggested a period increase. Finally, \citet{sing2024} measured a dynamical mass of $1.170 \pm 0.043$\,$M_{\rm J}$ from planetary atmospheric Doppler shifts.

From our analysis, we found $\Delta \mathrm{AIC} = -16.82$ and $\Delta \mathrm{BIC} = -13.84$, meaning a statistically significant \jkt{favor towards}{preference for} the quadratic model. The median quadratic model provides a secular increase in the orbital period with ${dP_{\rm orb}}/{dt} = 15.1 \pm 0.8 {\rm ~ms~yr^{-1}}$. Since the stellar spin is faster than the planetary orbit (P$_{\rm orb}$ = 1.27 d; Table \ref{tab:exoplanet_systems} and P$_{\rm rot}$ = 1.15 d; Table \ref{tab:mesa_results}), the orbital growth might be a viable explanation for this result. Lastly, we could not provide a constraint on $Q_\star^{\prime}$, as both limits of $\dot{P}$ resulted in a positive value.

Our theoretical models for WASP-121 indicate that the most efficient tidal mechanism is likely to be inertial waves, with a typical value $Q^{'}_\mathrm{IW}\approx 6.6\times 10^6$ at 1.5 Gyr. Gravity waves are predicted to be less effective, with $Q^{'}_\mathrm{IGW}\approx 1.48\times 10^{10}$ at the same age, and wave breaking is not predicted in the star. The star rotates with a period of $P_\mathrm{rot}\approx 1.13$ d \citep{bourrier2020}, indicating that inertial waves will be excited in the convective envelope and we also expect outward migration due to these waves, since $\mathrm{P}_\mathrm{rot} < \mathrm{P}_\mathrm{orb}$.

\begin{table*}
\centering
\caption{Stellar properties related to the efficient IW and IGW dissipation for the systems in our sample.}
\label{tab:mesa_results}
\renewcommand{\arraystretch}{1.5}
\begin{tabular}{lcccccccc}
\hline
System & Age & $P_{\mathrm{rot}}$\textsuperscript{\hyperref[fn:bb]{b}} & $v \sin i$ & $Q^{'}_{\mathrm{IW}}$ & $Q^{'}_{\mathrm{IGW}}$ & $M_\text{crit}$\textsuperscript{\hyperref[fn:cc]{c}} & Wave breaking? & Reference\textsuperscript{\hyperref[fn:aa]{a}}\\
 & (Gyr) & (days) & (km s$^{-1}$) &  &  & ($M_{\rm J}$) & \\
\hline
CoRoT-2 & 2.7 & 4.48 & $11.85$ & - & $1.94\times 10^6$ & 1.4 & Yes & 1 \\
HAT-P-23 & 4 & 7.01 & $8.10$ & - & $6.7\times 10^5$ & - &  No & 2 \\
HATS-18 & 4.4 & 9.8 & 6.23 & - & $1.2\times 10^5$ & $\sim 1$ & Yes & 3 \\
KELT-9 & 0.3 & 0.79 & 111.40  & $\sim 10^{10}$ & $\sim 10^{12}$ & - & No & 4 \\
KELT-16 & 2 & 9.05 & 7.6 & - & $1.5\times 10^6$ & - & No & 5\\
Qatar-1 & 11.6 & 28 & 1.70  & - & $1.85\times 10^5$ & 0.5 & Yes &  1\\
Qatar-4 & 0.15 & 6.0 & 7.1 & - & $1.63\times 10^6$ & $\gtrsim 10^3$ & No & 6 \\
TOI-1937A & 3.6  & 6.5 & - & - & $2.33\times 10^5$ & 0.75 & Yes & 7 \\
TOI-2109 & 1.77 & 1.14 & 81.2 & $6.9\times 10^6$ & $1.2\times 10^8$ & - & No & 8\\
TrES-1 & 3.7  & 31.4 & 1.3 & - & $2.57\times 10^6$ & 15.4 & No & 1 \\
WASP-3 & 2.1 & 5.15 & 13.4 & - & $3.8\times 10^7$ & - & No & 1\\
WASP-12 & 3.05 & 37.4 & 2.2 & - & $1 \times 10^5$ & $0.1$ & Yes & 1 \\
WASP-19 & 6 & 10.5 & 4.0 & - & $6\times 10^4$ & 1.0 & Yes, for older stars & 1\\
WASP-46 & 9.6 & 16.0 & 1.9 & - & $2.58\times 10^5$ & 1.7 & Yes, marginally & 1 \\
WASP-87 & ? & 30 & 9.9 & - & $4.5\times 10^6$ & - & No & 9 \\
WASP-103 & 1.7? & 6.85 & 10.6 & - & $1.8\times 10^6$ & - & No & 1\\
WASP-114 & 4.3  & 6.85 & 6.4 & - & $4.8\times 10^3$ & $<$ 0.1 & Yes & 10 \\
WASP-121 & 1.5 & 1.13 & 13.5 & $6.6\times 10^6$ & $1.48\times10^{10}$ & - & No & 11 \\
WASP-122 & 1.44 & $\sim$30 & 3.3 & - & $1.1\times 10^7$ & - &  No & 12\\
WASP-167 & 1.3& 1.88 & - & $6.5\times 10^9$ & $4\times 10^{14}$ & - & No & 13 \\
\hline
\end{tabular}
\parbox{\linewidth}{\footnotesize{
\label{fn:aa}$^\text{a}$ Reference for $P_\text{rot}$ or $v\sin{i}$.\par
\label{fn:bb}$^\text{b}$ For the stars whose $P_\mathrm{rot}$ has not been provided by previous literature sources, we used the simple formula of $P_\mathrm{rot} = 2\pi R_\star / v\sin{i}$.\par
\label{fn:cc}$^\text{c}$ We indicate all stars with convective cores using "-" since wave breaking is not predicted in them.}}
\begin{flushleft}
    {\footnotesize 
    \textbf{References:}
    1. \cite{bonomo2017}, 
    2. \cite{salisbury2021},
    3. \cite{penev2016},
    4. \cite{gaudi2017},
    5. \cite{oberst2017},
    6. \cite{alsubai2017},
    7. \cite{jankowski2025},
    8. \cite{wong2021},
    9. \cite{addison2016},
    10. \cite{addison2016},
    11. \cite{delrez2016},
    12. \cite{turner2016},
    13. \cite{temple2017}.
    }
    \end{flushleft}
\end{table*}

\subsubsection{WASP-122 (KELT-14)}

WASP-122\,b (or KELT-14\,b) is an important candidate for displaying orbital decay because it orbits a G2 star near the main sequence turnoff \citep{rodriguez2016}. Previous studies have noted its potential for orbital decay (\citealt{patra2020}) and compiled transit timings from the literature and TESS observations (\citealt{shan2023,maciejewski2022a}), but none found evidence for deviations from a linear ephemeris. Rossiter-McLaughlin observations indicate that the orbit is circular and well-aligned \citep{stangret2024}. 

From our TTV analysis, we found $\Delta {\rm AIC} = -1.13$ and $\Delta {\rm BIC} = 1.32$. Therefore, we conclude that there is no statistically significant preference for any model. Our median quadratic model suggests that the 99.7\% confidence-level lower limit of $a_1^{\rm lim} = -2.822\times10^{-10}~\text{d cycle}^{-1}$ corresponds to $Q_\star^{\prime} > (4.6 \pm 0.5) \times 10^4$.

Our theoretical models of WASP-122 predict that gravity waves (in the fully-damped regime) can provide $Q^\prime_\mathrm{IGW}\approx 1.1\times 10^7$ at 1.44 Gyr. The planet is unlikely to cause these waves to break though magnetic wave conversion is a possibility to justify this fully damped regime, particularly if the star is near the end of the MS or older \citep{duguid2024}. The star probably rotates far too slowly for inertial waves to be excited. Lastly, it should be noted that the estimated age from \citet{turner2016} is approximately 5 Gyr older than what we present for efficient IGW dissipation in the fully-damped regime, though the prior age constraint may be unreliable.

\subsubsection{WASP-167 (KELT-13)}

WASP-167\,b is a neglected hot-Jupiter in the literature although it has been over eight years since its discovery \citep{temple2017}. It orbits an F1-type main-sequence star, which is a $\gamma$-Doradus pulsator \citep{kalman2024}. \citet{kokori2023} and \citet{kokori2022}, \citet{ivshinawinn2022}, and \citet{kokori2023} revised its ephemeris information but did not note any departure from the linear ephemerides.

From our timing analysis, we found an increasing period with a rate of {$11.8 \pm 4.1~\text{ms yr}^{-1}$}. However, since the corresponding $\Delta {\rm AIC}$ and $\Delta {\rm BIC}$ do not exceed the threshold of 10, these findings are not statistically significant. 
Additionally, by adopting the 99.7\% confidence level upper limit for $dP/dt$, we provide a lower limit on $Q_{\star}^{\prime} > (10.9 \pm 1.1) \times 10^7$.

Our theoretical models of WASP-167 indicate that gravity waves are probably inefficient, with $Q^{'}_\mathrm{IGW}\gtrsim 10^{14}$ at 1.3 Gyr. On the other hand, the rapid rotation of the star, $P_\mathrm{rot}=1.88$ d \citep{temple2017} implies that inertial waves are likely to be excited in the stellar convective envelope. Since $P_\mathrm{rot}<P_\mathrm{orb}$, it is likely that these waves would drive outward planetary migration, though our models predict $Q^{'}_\mathrm{IW}\approx 6.5\times 10^9$ at 1.3 Gyr as a representative value for the dissipation of these waves. Hence, tidally-driven orbital evolution is predicted to be very slow and unlikely to be detected according to our current theoretical understanding.

\section{Discussion}
\label{sec:discussion}

\subsection{Apsidal motion scenario}
\label{subsec:apsidalmotion}

Apart from the constant period or secularly changing orbital period model, tidally induced apsidal precession is an alternative explanation for such TTV signals \citep[e.g.][]{barker2024}. For the systems with even a slight eccentric orbit, tidally induced apsidal precession produces quasi-periodic variations in the transit and occultation times. This mechanism has been discussed for several HJs in previous studies (e.g., KELT-9\,b; \citeauthor{harre2023}~\citeyear{harre2023}, WASP-19\,b; \citeauthor{biswas2024}~\citeyear{biswas2024}). Being motivated by these previous attempts, we followed the formalism from \citet{harre2023} to test the apsidal motion hypothesis for nine HJs in our sample. For the rest of the sample, previous works have either ruled out the apsidal motion scenario or provided a circular orbit followed by subsequent measurements. Briefly, the apsidal precession model assumes a non-zero eccentricity and is defined as
\begin{equation}
    t_{\rm tra}(N) = t_0 + E P_s - \frac{e P_a}{\pi} \cos \omega(E),
\end{equation}
where
\begin{equation}
    \omega(E) = \omega_0 + \frac{d\omega}{dE}E,
\end{equation}
and
\begin{equation}
    P_s = P_a \left(1 - \frac{1}{2\pi}\frac{d\omega}{dE}\right).
\end{equation}
Here, $P_s$ is the sidereal period, $P_a$ is the anomalistic period and $\omega$ is the argument of periastron. We present our results with a $\Delta \rm BIC$ compared to the constant period model in Table \ref{tab:ttv_results_aps}. That is, we replaced the quadratic model with apsidal motion model in Eq. \ref{eq:bic}.  Despite the physical plausibility of the mechanism, we have found no statistically significant evidence for apsidal precession. For all fits, $\Delta \rm BIC > 140$ and is consistent with a circular orbit at 99.7\% confidence level.

Several of the previously reported eccentricities or findings on the apsidal motion models in the literature are themselves subject to significant uncertainties and, in some cases, contradictory results. Small eccentricities inferred from RV measurements are particularly vulnerable to degeneracies between eccentricity and argument of periastron, as well as to the influence of stellar jitter and correlated noise \citep{shen2008}. It is therefore possible that at least some of the systems in our sample are in fact on nearly circular orbits, possibly as a result of tidal circularization \citep{valsecchi2014}, in which case apsidal motion would not produce a measurable transit timing signal. Tidal circularization is usually thought to be primarily produced by dissipation inside the planet rather than the star \citep[e.g.][]{ogilvie2014,Lazovik2024}. \citet{gillon2010} reported a statistically significant eccentricity value of $0.0143^{+0.0077}_{-0.0076}$ for CoRoT-2\,b's orbit by constraining the \jkt{cosine projection of eccentricity}{quantity} $e\cos\omega$ with the help of \emph{Spitzer} occultation observations. Therefore, an apsidal motion can be expected, but the relatively low precision of the \emph{Spitzer} occultation timings prevents us from investigating this possibility in detail. For WASP-3\,b, the orbit had been reported to be eccentric based on occultation \citep{rostron2014} and RV observations \citep{knutson2014,bonomo2017}. For HAT-P-23\,b, \emph{Spitzer} secondary-eclipse measurements suggest a circular orbit \citep{orourke2014}, indicating past tidal circularization and alignment \citep[e.g.][]{valsecchi2014}; however, this is contradicted by the result of \citet{stassun2017}, who found \(e = 0.11 \pm 0.04\). Although \cite{gaudi2017} suggested a circular orbit for KELT-9\,b's orbit, \cite{harre2023} showed that the apsidal motion model better fit the data with $\Delta {\rm BIC} = -13.28$, so we considered this case as well. 
In order to test this mechanism for all systems, we encourage further occultation and radial velocity observations with adequate precision for eccentricity measurements and mid-occultation timings, which will vary in the opposite phase of the mid-transit timings in the presence of apsidal motion.


\subsection{Comparison to the previous theoretical results}

For the hosts of exoplanets on tight ($a < 0.05 {\rm ~au}$) and eccentric orbits, \cite{bonomo2017} proposed $6< \log_{10}Q_\star^{\prime} < 7$ by comparing the circularization times of HJs and their host stars. This inferred interval is in agreement with some of our systems for which strictly circular orbits are not explicitly stated (Table \ref{tab:ttv_results_aps}). To specify, either observational limits or theoretical expectations lie within the specified range for CoRoT-2, HAT-P-23, HATS-18, KELT-9, TOI-2109, WASP-19, WASP-114 and WASP-167. However, this agreement is somewhat puzzling for these systems, since wave breaking is also expected to trigger efficient IGW dissipation for some of the host stars when the host star is approaching the terminal stage of the main sequence phase or has already entered the subgiant phase \citep{barker2020}. These two expectations do not appear to be fully consistent with each other, as the circularization timescale $\tau_c$ should be smaller than the inspiral timescale $\tau_a$, otherwise we would not be able to see these objects in their current orbits \citep{ogilvie2014}. This might be the result of tidal dissipation inside the planet dominantly driving the circularization, instead of dissipation inside the star. \par

One of the most recent estimates was provided from \cite{hamerschlaufman2019}, where they found that HJs with a mass of $1 ~M_{\rm J} < M_{\rm p} < 2 ~M_{\rm J}$ and an orbital period of $2 {\rm ~d}< P_{\rm orb} <5~{\rm d}$ should satisfy $Q_\star^{\prime} \lesssim 10^7$. This limit appears as the upper bound of the proposed value from \cite{bonomo2017}. If we also extend this estimate to our sample of HJs with $P_{\rm} < 2$~d, it can be concluded that the majority of our results estimates from timing analysis and theory are in approximate agreement with this value. Only the inferred $Q^{'}_{\rm IW}$ for KELT-9 and WASP-167 exhibits a weaker dissipation by at least 2 orders of magnitude. 

\subsection{A Sanity Check for Tidal Stability}

\begin{figure}
    \centering
    \includegraphics[width=0.45\textwidth]{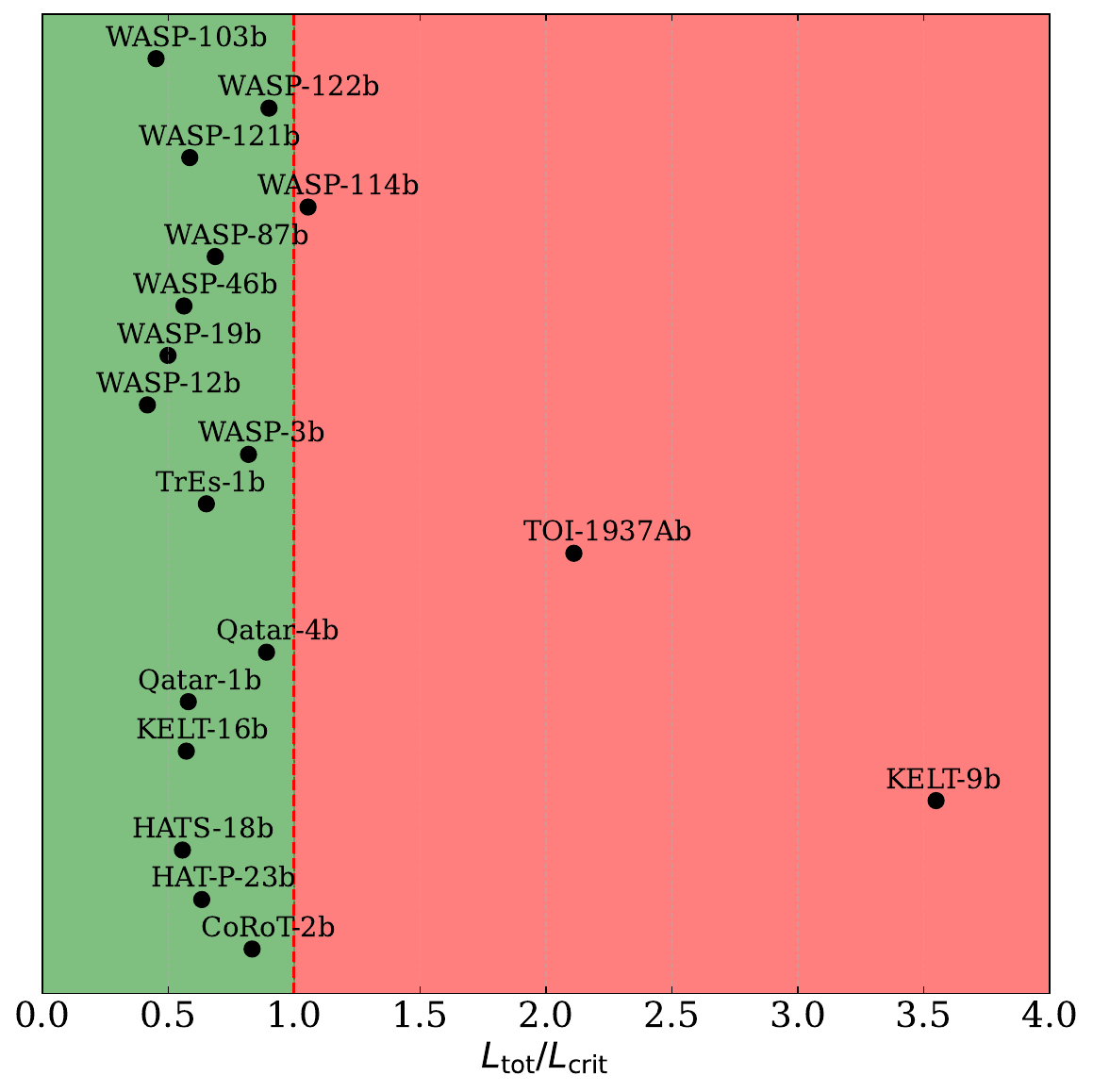}
    \caption{The ratio of the total angular momentum $L_{\rm tot}$ to the critical angular momentum $L_{\rm crit}$ for the studied HJ systems. The green region corresponds to $L_{\rm tot}/L_{\rm crit} < 1$, where the system is ``Darwin unstable", while the red region corresponds to the $L_{\rm tot}/L_{\rm crit} >1$, where the stability of the system depends on the relationship between $L_{\rm orb}$ and $L_{\rm spin}$.}
    \label{fig:ltot_lcrit}
\end{figure}

Following \cite{hut1980} (see also \citealt{levrard2009} and \citealt{matsumura2010}), we also performed an additional check of the ``stability'' of the systems in our sample.
If the total angular momentum ($L_{\rm tot}$; the sum of the rotational $L_{\rm spin}$ and orbital $L_{\rm orb}$ components) of a system is smaller than a critical value ($L_{\rm crit}$), i.e., $L_{\rm tot} < L_{\rm crit}$, then we say that the system is ``Darwin unstable'' and the HJ will spiral into the star to reach the Roche limit. On the other hand, if $L_{\rm tot} > L_{\rm crit}$, there are two outcomes: (1) $L_{\rm orb} <3 L_{\rm spin}$ (or $L_{\rm orb}/ L_{\rm spin} <$ 3) then the HJ will spiral into its star to reach the Roche limit, or (2) $L_{\rm orb} > 3L_{\rm spin}$ and the orbit is ``stable'', such that the planet will not spiral inward to reach the Roche limit. To infer which outcome we should expect in our systems if the total angular momentum is conserved, we first calculated the total angular momentum $L_{\rm tot}$ and the critical angular momentum $L_{\rm crit}$, which are defined by
\begin{equation}
    L_{\rm tot} = \underbrace{C_\star \omega_\star}_{=L_\mathrm{spin}} + \underbrace{\frac{M_{\rm p}M_{\rm \star}}{\sqrt{M_{\rm p}+M_\star}}\sqrt{Ga(1-e^2)}}_{=L_\mathrm{orb}},
\end{equation}
and
\begin{equation}L_{\text{crit}} = 4 \left[ 
\frac{G^2}{27} \, \frac{M_\star^3 \, M_{\rm p}^3}{M_\star + M_{\rm p}} \, (C_\star + C_{\rm p})
\right]^{\tfrac{1}{4}},
\end{equation}
\noindent where $G$ is the gravitational constant, $C_\star = k_\star M_\star R_\star^2$ is the moment of inertia of the star, where $M_\star$ is the stellar mass, with $C_{\rm p}$ and $M_{\rm p}$ for the planet defined similarly, $a$ is the semi-major axis and $e$ is the eccentricity. We adopted $k_\star = 0.06$ for the host stars \citep[e.g.][]{claretgimenez1989} and $k_p= 0.26$ for the HJs in our sample \citep[e.g.][]{gu2003,lanza2020}.
The results are displayed in Fig.~\ref{fig:ltot_lcrit}. Only three HJ systems (KELT-9, TOI-1937A and WASP-114\,b) have $L_{\rm tot} > L_{\rm crit}$, such that there is currently enough angular momentum in the system for a stable tidal equilibrium state to exist in the absence of angular momentum loss due to stellar magnetic braking. For the remaining 17 systems, each one is predicted to have insufficient angular momentum for a stable equilibrium state to exist, and hence these planets are expected to spiral into their stars to destruction. However, the red region in Figure \ref{fig:ltot_lcrit} does not necessarily imply that the orbit is stable, so we also checked which portion of the parameter space these 3 HJ systems occupy according to the relation between $L_{\rm orb}$ and $L_{\rm spin}$. For KELT-9\,b, TOI-1937A\,b and WASP-114\,b, we found $L_{\rm orb}/L_{\rm spin} = 0.11$, $L_{\rm orb}/L_{\rm spin} = 0.08$ and $L_{\rm orb}/L_{\rm spin} = 0.07$, respectively. Therefore, we conclude that all the HJs in our sample are ``Darwin unstable'', so we expect these planets to be engulfed after they have spiralled inside the Roche radius of their host-stars.

It should be remembered that angular momentum loss from the star-planet system, such as by stellar magnetic braking, will preclude an ultimate tidal equilibrium state in any of these systems even if the above criterion would have otherwise predicted them to be ``stable'' \citep[e.g.][]{BO2009,Damiani2015,B2025}. Nevertheless, this check is useful for interpretation for cases where this angular momentum loss is expected to be weak, such as in the more massive stars in our sample, in particular KELT-9.

\section{Conclusion}
\label{sec:conclusion}

In this work, we first performed a target selection process to select the most suitable targets to investigate the orbital decay of HJs. We selected 20 systems with HJs and presented 42 new ground-based observations for 12 of these. 
With the help of measurements from ETD, TESS and the literature, we have investigated the likelihood of tidally-driven orbital decay in each of these 20 HJ systems by exploring the significance of an orbital decay model in each system's TTV diagram. We fitted a linear (constant orbital period) model to each one and an apsidal motion model to those with eccentric orbits. We complemented this analysis with theoretical models of the stellar tidal dissipation.

We were able to confirm and update the parameters for the orbital decay of WASP-12\,b, which has been \jkt{well-}{extensively} studied for over a decade. We have also confirmed \jkt{previous arguments supporting that}{that} either WASP-12 should be in a subgiant phase to satisfy the observationally-inferred $Q_*^{\prime} \approx 10^5$ in theoretical models, 
or that magnetic wave conversion is in operation if WASP-12 is a main-sequence star \citep{duguid2024}. 

Additionally, our timing analysis has revealed statistically-favored orbital period changes for TrES-1\,b and WASP-121\,b. We have found that the orbital period is decreasing at a rate of $-14.9\pm0.6$ ~ms yr$^{-1}$ for TrES-1\,b and increasing at a rate of $15.1\pm0.8$ ~ms yr$^{-1}$ for WASP-121\,b. TrES-1\,b's decay implies very efficient stellar tidal dissipation with $Q_*^{\prime} = 5.7 \times 10^2$, which is incompatible with our theoretical models. For this system, we do not predict gravity waves to break in the radiative core of the host star, and $Q^\prime_\star$ is predicted to be \jkt{4}{four} orders of magnitude larger when gravity waves are fully damped. Also, when we compare this inferred $Q_*^{\prime}$ with cases of WASP-12\,b (\S \ref{subsub:wasp12}) and WASP-4\,b \citep{basturk2025}, we \jkt{determined}{find} that the outcome of TTV analysis does not agree with the expectations for similar systems. On the contrary, inertial waves can provide efficient tidal dissipation in the convective envelope of WASP-121, and are expected to cause orbital growth, consistent with the results of our timing analysis.

Apart from these three systems, our timing analysis indicates no significant evidence in favor of either a decaying or growing orbit in the remaining systems in our sample. Nevertheless, for almost half of the planets in our list, wave breaking of gravity waves in the stellar radiative core is expected and should lead to efficient gravity wave dissipation. For those in which wave breaking is not expected, such as in hotter stars with convective cores, magnetic wave conversion is an alternative explanation to provide efficient damping of gravity waves and the resulting $Q_*^{\prime}$ \citep{duguid2024}. Another argument to motivate monitoring these objects further is that all of these systems are ``Darwin unstable'', meaning that eventually the HJ will migrate towards the Roche limit of its host star. Therefore, we believe that our work could motivate future follow-up studies of these systems. At least half of the systems in our sample will be observed by TESS in 2026 or 2027. In addition, the PLATO mission is expected to be launched before the end of 2026 \citep{rauer2025}. WASP-121, WASP-122, TOI-1937A, and other systems with potential but not in our sample lie within the LOPS2 field \citep{nascimbeniPLATO}. Thus, we expect further observational constraints to be imposed on the tidal quality parameters of the hosts of these systems and therefore to obtain a better understanding of tidal interactions in the coming years.

\begin{acknowledgments}

We gratefully acknowledge the support of the Scientific and Technological Research Council of T{\"u}rkiye (TÜBİTAK) with project
123F293.
In this study, the photometric observations obtained within the scope of project numbered 12CT100-378 and 24BT100-2148, and carried out using the TUG100 telescope at the TUG site of the T{\"u}rkiye National Observatories have been utilised; we express our gratitude for the support provided by the T{\"u}rkiye National Observatories, the observation team and all its staff.
Some of the data in this study were obtained with the T80 telescope at the Ankara University Astronomy and Space Sciences Research and Application Center (Kreiken Observatory) with project numbers 23B.T80.06 and 25BT80.01.
We thank all the observers who report their observations to ETD and AXA open data bases.
This work includes data collected with the TESS mission, obtained from the MAST data archive at the Space Telescope Science Institute (STScI). Funding for the TESS mission is provided by the NASA Explorer Program. STScI is operated by the Association of Universities for Research in Astronomy, Inc., under NASA contract NAS 5–26555. The authors acknowledge the use of public TESS data from pipelines at the TESS Science Office and at the TESS Science Processing Operations Center. Resources supporting this work were provided by the NASA High-End Computing (HEC) Program through the NASA Advanced Supercomputing (NAS) Division at Ames Research Center for the production of the SPOC data products. 
The authors acknowledge the use of the 2.2\,m Telescope at the ESO Observatory in La Silla (Chile) and the use of the GROND camera, which was built by the high-energy group of MPE in collaboration with the LSW Tautenburg and ESO.
We acknowledge the use of the Danish 1.54\,m telescope at the ESO La Silla Observatory, within the framework of the MiNDSTEp consortium.
The authors acknowledge the use of the Cassini 1.52\,m Telescope operated by INAF-OAS (Astrophysics and Space Science Observatory of Bologna) in Loiano. 
The authors thank Ivan Bruni and Roberto Gualandi for their technical assistance at the Loiano Observatory.
ACK gratefully acknowledges the financial support of the TÜBİTAK 2210 National Graduate Scholarship no. 1649B022401264, UK
Science and Technology Facilities Council (STFC) and the Faculty
of Natural Sciences, Keele University in the form of a PhD studentship.
AJB was supported by STFC grants ST/W000873/1 and UKRI1179. L.M. acknowledges support from the MIUR-PRIN project no. 2022J4H55R.
JS acknowledges support from STFC under grant number ST/Y002563/1.
SA acknowledges the funding from the Scientific Research Projects Coordination Unit of Istanbul University with the project number: FBA-2022-39121. IST60 telescope and its equipment are supported by the Republic of Turkey Ministry of Development (2016K12137) and Istanbul University with the project numbers BAP-3685, FBG-2017- 23943.
A\"O acknowledges T\"UB{\.I}TAK B{\.I}DEB for the support in terms of 2210 program.
LM acknowledges financial contribution from PRIN MUR 2022 project 2022J4H55R.
The work presented here is supported by the Carlsberg Foundation, grant CF25-0040.
RJFJ acknowledges the support provided by the GEMINI/ANID project under grant number 32240028, by ANID’s Millennium Science Initiative through grant ICN12\_009, awarded to the Millennium Institute of Astrophysics (MAS), and by ANID’s Basal project FB210003.

\end{acknowledgments}

\facilities{TRG\"OZ(TUG100), AUKR(T80), TRG\"OZ(ATA050), IUGUAM (IST60), El Sauce (ODK20, CDK20), La Silla (DK154, GROND, NTT), CAHA, Loiano (BFOSC)}

\software{astropy \citep{astropy2013,astropy2018,astropy2022}, AstroImageJ \citep{astroimage2017}, EXOFAST \citep{eastman2013}, matplotlib \citep{matplotlib}, MESA \citep{Paxton2011,Paxton2013,Paxton2015,Paxton2018,Paxton2019,Jermyn2023},  numpy \citep{numpy}, pandas \citep{reback2020pandas}, pymc \citep{pymc2023}, scipy \citep{2020SciPy-NMeth}.}


\appendix
\restartappendixnumbering
\section{Additional tables}
We provide additional information on the dataset of light curves that we used throughout the study. For Table \ref{tab:etd_observations},  \ref{tab:lit_adopted}, \ref{tab:lit_observations} and \ref{tab:tess_observations}, the full table is available at the CDS table.
\begin{table} 
\centering 
\caption{Analyzed TESS Sectors for systems in our sample.} \label{tab:tess_sectors} 
\begin{tabular}{ll} 
\hline \textbf{System} & \textbf{TESS Sector} \\ 
\hline CoRoT-2 & 54, 81 \\ 
HAT-P-23 & 54, 55, 81 \\ 
HATS-18 & 10, 36, 63, 90 \\ 
KELT-9 & 14, 15, 41, 55, 75, 82 \\ 
KELT-16 & 15, 41, 55\\ 
Qatar-1 & 17, 24, 41, 48, 51, 55, 56, 57, 59, \\ & 75, 76, 77, 78, 82, 83, 84, 85, 86 \\ 
Qatar-4 & 57, 84 \\ 
TOI-2109 & 52, 79 \\ 
TOI-1937A & 34, 35, 36, 61, 62, 88, 89 \\ 
TrEs-1 & 14, 41, 53, 54, 74, 80 \\
WASP-3 & 26, 40, 53, 54, 74, 80, 81 \\
WASP-12 & 20, 43, 44, 45, 71, 72 \\
WASP-19 & 9, 36, 62, 63, 89, 90 \\ 
WASP-46 & 1, 27, 67 \\ 
WASP-87 & 10, 11, 37 \\ 
WASP-103 & -\\ 
WASP-114 & 55, 82 \\
WASP-121 & 7, 33, 34, 61, 87, 88\\
WASP-122 & 7, 33, 34, 61, 87, 88 \\
WASP-167 & 10, 37, 64\\ 
\hline 
\end{tabular} 
\end{table}

\begin{table}
\centering
\caption{Number of light curves analysed and minimum adopted from the literature sources for each planetary system in our sample. L$_\text{1}$ is the number of adopted minima times from the literature sources, L$_\text{2}$ is the light curve number originated from literature and analysed using AIJ and {\sc exofast}, and T is the total number of minimum times.}
\begin{tabular}{lcccccc}
\hline
Planet & ETD & L$_\text{1}$ & L$_\text{2}$ & TESS & Ours & T \\
\hline
CoRoT-2\,b & 38 & 33 & 86 & 18 & 0 & 175 \\
HAT-P-23\,b & 28 & 32 & 13 & 56 & 1 & 130 \\
HATS-18\,b &  4 & 2 & 28 & 105 & 2 & 141 \\
KELT-9\,b & 0 & 8 & 23 & 94 & 0 & 125 \\
KELT-16\,b & 15 & 0 & 60 & 60 & 0 & 135 \\
Qatar-1\,b & 322 & 12 & 65 & 302 & 11 & 712 \\
Qatar-4\,b & 32	& 1 & 1 & 24 & 1 & 59 \\
TOI-2109\,b & 0	& 11 & 23 & 47 & 3 & 84 \\
TOI-1937A\,b &  0 & 0 & 6 & 168	& 0	& 174\\
TrES-1\,b & 14 & 42 & 7 & 39 & 1 & 103 \\
WASP-3\,b & 68 & 19 & 56 & 79 & 3 & 212 \\
WASP-12\,b & 83 & 0 & 78 & 116 & 2 & 279 \\
WASP-19\,b & 50 & 31 & 36 & 174 & 10 & 301 \\
WASP-46\,b & 33 &  0 & 22 & 45 & 0 & 100 \\
WASP-87\,b & 3 & 0 & 7 & 37 & 0  & 47\\
WASP-103\,b & 8 & 82 & 0 & 0 & 2 & 92 \\
WASP-114\,b & 11 & 2 & 0 & 24 & 2 & 39 \\
WASP-121\,b & 9 & 17 & 0 & 105 & 4 & 135 \\
WASP-122\,b & 9 & 0 & 4 & 78 & 0 & 91 \\
WASP-167\,b & 5 & 0 & 5 & 32 & 0 & 42 \\
\hline
\textbf{Total}& 732 & 292 & 520& 1603 & 42 & 3188 \\
\hline
\end{tabular}
\label{tab:number_lcs}
\end{table}

\begin{table}
\centering
\caption{Same as Table \ref{tab:number_lcs} but after selection criterion applied on the dataset.}
\begin{tabular}{lcccccc}
\hline
Planet & ETD & L$_\text{1}$ & L$_\text{2}$ & TESS & Ours & T \\
\hline
CoRoT-2\,b & 30 & 33 & 80 & 14 & 0 & 153 \\
HAT-P-23\,b & 25 & 32 & 13 & 55 & 1 & 126 \\
HATS-18\,b & 4 & 2 & 28 & 102 & 2 & 138 \\
KELT-9\,b & 0 & 8 & 23 & 94 & 0 & 125 \\
KELT-16\,b & 7 & 0 & 36 & 53 & 0 & 96 \\
Qatar-1\,b & 304 & 12 & 65 & 302 & 10 & 693 \\
Qatar-4\,b & 29 & 1 & 1 & 23 & 1 & 55 \\
TOI-2109\,b & 0 & 11 & 15 & 45 & 0 & 71 \\
TOI-1937A\,b & 0 & 0 & 5 & 156 & 0 & 161 \\
TrEs-1\,b & 10 & 42 & 6 & 37 & 1 & 96 \\
WASP-3\,b & 53 & 19 & 52 & 77 & 3 & 204 \\
WASP-12\,b & 49 & 0 & 69 & 113 & 2 & 233 \\
WASP-19\,b & 48 & 31 & 33 & 171 & 7 & 290 \\
WASP-46\,b & 26 & 0 & 17 & 45 & 0 & 88 \\
WASP-87\,b & 0 & 0 & 3 & 21 & 0 &24 \\
WASP-103\,b & 8 & 82 & 0 & 0 & 2 & 92 \\
WASP-114\,b & 9 & 2 & 0 & 20 & 2 & 33 \\
WASP-121\,b & 3 & 17 & 0 & 103 & 4 & 127 \\
WASP-122\,b & 4 & 0 & 4 & 77 & 0 & 85 \\
WASP-167\,b & 3 & 0 & 5 & 28 & 0 & 36 \\
\hline
\textbf{Total}& 612 & 292 & 455 & 1536 & 35 & 2930 \\
\hline
\end{tabular}
\label{tab:lcs_after_pnr_beta}
\end{table}

\begin{table*}
\centering
\caption{Mid-transit times we derived from the light curves in the Exoplanet Transit Database (ETD) ($T_{0,\rm cal}$), their uncertainties ($\sigma$), mid-transit times reported by the observers of ETD from the same light curves $T_{0,{\rm ref}}$.}
\label{tab:etd_observations}
\begin{tabular}{lccccccccl}
\hline
System & Observer & Filter & $T_{0,{\rm ref}}$ & $T_{0,{\rm cal}}$ & $\sigma$ & omit? & $\beta$ & PNR & Comments \\
\hline
CoRoT-2 & Yves Jongen & clear & 2460559.386569 & 2460559.387443 & 0.000362 & 0 & 1.924 & 1.546 & \\
CoRoT-2 & Yves Jongen & clear & 2460067.861817 & 2460067.861979 & 0.000254 & 0 & 2.291 & 1.252 & \\
CoRoT-2 & Yves Jongen & clear & 2459431.668172 & 2459431.668660 & 0.000292 & 0 & 2.141 & 1.372 & \\
CoRoT-2 & Yves Jongen & clear & 2459412.495652 & 2459412.496588 & 0.000460 & 0 & 1.444 & 2.103 & \\
... & ... & ... & ...& ...& ...& ...& ... & ... & ... \\
\hline
\end{tabular}
\end{table*}

\begin{table*}[htbp]
\centering
\caption{Mid-transit times we adopted from previous works. Note that we provide the bibcode of the study in the CDS version of this table.}
\label{tab:lit_adopted}
\begin{tabular}{llcc}
\hline
System & Source & $T_{0,{\rm ref}}$ & $\sigma_{\rm ref}$ \\
\hline
CoRoT-2 & \cite{bouchy2008} & 2454237.536 & 0.00014 \\
CoRoT-2 & \cite{rauer2010,adams2024} & 2454624.481 & 0.0016 \\
CoRoT-2 &  \cite{rauer2010,adams2024} & 2454676.762 & 0.0008 \\
CoRoT-2 &  \cite{rauer2010,adams2024} & 2454678.502 & 0.0012 \\
CoRoT-2 &  \cite{rauer2010,adams2024} & 2454683.741 & 0.0014 \\
... & ... & ... & ...  \\
\hline
\end{tabular}
\end{table*}

\begin{table*}
\centering
\caption{Mid-transit times we derived from the light curves in the previous studies ($T_{0,\rm cal}$), their uncertainties ($\sigma_{\rm cal}$), mid-transit times reported in the corresponding study ($T_{0,{\rm ref}}$) and their uncertainties ($\sigma_{\rm ref}$). Note that we provide the bibcode of the study in the CDS version of this table.}
\label{tab:lit_observations}
\begin{tabular}{llccccccl}
\hline
System & Source & $T_{0,{\rm ref}}$ & $\sigma_{\rm ref}$ & $T_{0,\rm cal}$ & $\sigma_{\rm cal}$ & $\beta$ & PNR & Comments \\
\hline
CoRoT-2 & \cite{adams2024} & 2454237.534500 & 0.0017 & 2454237.535189 & 0.000265 & 2.972 & 0.273 & Beta criteria \\
CoRoT-2 & \cite{adams2024} & 2454239.278590 & 0.00029 & 2454239.278895 & 0.000251 & 1.628 & 3.377 & \\
CoRoT-2 & \cite{adams2024} & 2454241.022200 & 0.00024 & 2454241.022116 & 0.000271 & 1.525 & 3.382 & \\
CoRoT-2 & \cite{adams2024} & 2454242.765440 & 0.00025 & 2454242.765626 & 0.000161 & 1.539 & 3.119 & \\
CoRoT-2 & \cite{adams2024} & 2454244.508370 & 0.00025 & 2454244.508497 & 0.000193 & 1.458 & 2.924 & \\
... & ... & ... & ... & ... & ... & ... & ... & ... \\
\hline
\end{tabular}
\end{table*}

\begin{table*}
\centering
\caption{Mid-transit times we derived from the TESS light curves ($T_{0,\rm cal}$), their uncertainties ($\sigma_{\rm cal}$).}
\label{tab:tess_observations}
\begin{tabular}{lcccccc}
\hline
System & $T_{0,\rm cal}$ & $\sigma_{\rm cal}$ & $\beta$ & $PNR$ & Comments \\
\hline
CoRoT-2 & 2459771.551759 & 0.000381 & 1.944 & 2.361 & \\
CoRoT-2 & 2459773.294693 & 0.000399 & 2.211 & 2.334 & \\
CoRoT-2 & 2459783.753239 & 0.000428 & 2.528 & 2.589 & Beta criteria \\
CoRoT-2 & 2459785.495646 & 0.000404 & 1.959 & 2.365 & \\
CoRoT-2 & 2459787.238485 & 0.000421 & 2.134 & 2.410 & \\
... & ... & ... & ... & ... & ...\\
\hline
\end{tabular}
\end{table*}

\begin{table}
\centering
\caption{Adopted ephemeris information of our sample in this study.}
\renewcommand{\arraystretch}{1.3} 
\setlength{\tabcolsep}{2pt}
\begin{tabular}{lllc}
\hline
Planet & ~~~~$T_0$ (BJD$_{\rm TDB}$) & ~~~~~~$P_{\rm orb}$ (days) & Ref. \\
\hline
CoRoT-2\,b & 2457683.44158(16) & ~~~~~1.74299705(15) & 1 \\
HAT-P-23\,b & 2457742.573790(72) & ~~~~~1.212886397(74) & 1 \\
HATS-18\,b & 2458626.51102(45) & ~~~~~0.83784369(11) & 2 \\
KELT-9\,b & 2457095.68572(14) & ~~~~~1.4811235(11) & 3 \\
KELT-16\,b & 2458392.597691(78) & ~~~~~0.968992962(97) & 4 \\
Qatar-1\,b & 2456234.103218(60) & ~~~~~1.42002420(22) & 5 \\
Qatar-4\,b & 2457637.77361(46) & ~~~~~1.805356400(1) & 6\\
TOI-2109\,b & 2459378.459370(59) & ~~~~~0.672474140(28) & 7 \\
TOI-1937A\,b & 2459085.91023(7) & ~~~~~0.94667944(9) & - \\
TrES-1\,b & 2456822.891157(63) & ~~~~~3.030069476(72) & 1 \\
WASP-3\,b & 2454143.85112(24) & ~~~~~1.84683510(40) & 8 \\
WASP-12\,b & 2457010.512173(70) & ~~~~~1.091419108(55) & 9 \\
WASP-19\,b & 2456885.482836(37) & ~~~~~0.788839092(24) & 1 \\
WASP-46\,b & 2455392.31659(58) & ~~~~~1.43036763(93) & 10 \\
WASP-87\,b &  2458276.86087(15) & ~~~~~1.68279422(22) & 1 \\
WASP-103\,b & 2457308.324538(30)& ~~~~~0.925545386(56) & 1 \\
WASP-114\,b & 2457522.66047(24) & ~~~~~1.54877501(34) & 1 \\
WASP-121\,b & 2458119.72074(17) & ~~~~~1.27492504(15) & 12 \\
WASP-122\,b & 2456665.22401(21) & ~~~~~1.7100566(32) & 13 \\
WASP-167\,b & 2458117.02169(19) & ~~~~~2.02195933(33) & 1 \\
\hline
\multicolumn{4}{l}{
\parbox{\linewidth}{\footnotesize 
\textbf{References:} 
1. \cite{kokori2023}, 
2. \cite{southworth2022}, 
3. \cite{gaudi2017},
4. \cite{oberst2017},
5. \cite{collins2017qatar1},
6. \cite{alsubai2017},
7. \cite{wong2021},
8. \cite{bonomo2017},
9. \cite{ivshinawinn2022},
10. \cite{ciceri2016},
11. \cite{addison2016},
12. \cite{bourrier2020},
13. \cite{turner2016}.
}
}
\label{tab:ref_ephemeris}
\end{tabular}
\end{table}

\begin{table}
\centering
\caption{Median-fit results from our apsidal motion models for the HJs with $e>0$, compared with the linear model presented in Table \ref{tab:ttv_results}.}
\label{tab:ttv_results_aps}
\renewcommand{\arraystretch}{1.1} 
\begin{tabular}{lcccccc}
\hline
System & $\Delta \mathrm{BIC}$ & $e$ & $\omega~(\text{rad})$ & $\omega dE$\\
 & & $(\times 10^{-3})$ & & ($\times 10^{-4~~\circ}/ \text{cycle})$ \\
\hline
CoRoT-2 &  3021.216 & $1.01^{+0.19}_{-0.31}$ & $1.96^{+0.11}_{-0.14}$ & $17.47^{+2.59}_{-2.70}$ \\
HAT-P-23 & 688.66 & $1.49^{+0.967}_{-1.18}$ & $3.14^{+1.78}_{-1.79}$ & $5.81_{-5.63}^{+5.75}$\\
HATS-18 & 457.149 & $1.106^{+0.770}_{-1.382}$ & $2.70^{+1.78}_{-2.32}$ & $3.28^{+4.88}_{-5.98}$\\
KELT-9 & 215.01 & $0.301^{+0.148}_{-0.301}$ & $2.90^{+0.64}_{-0.82}$ & $15.3^{+3.03}_{-3.04}$ \\
WASP-3 & 472.06 & $0.55^{+0.30}_{-0.56}$ & $1.73^{+0.940}_{-1.27}$ & $372.7^{+101.2}_{-105.5}$ & \\
WASP-19 & > 5000 & $3.4991^{+0.0014}_{-0.0006}$ & $3.28^{+3.07}_{-3.00}$ & $1182.413^{+7.192}_{-6.799}$ \\
TOI-2109 & 187.26 & $1.70^{+1.09}_{-1.20}$ & $3.15^{+1.99}_{-2.00}$ & $5.74^{+5.69}_{-5.75}$ \\
WASP-114 & 146.95 & $1.392^{+0.904}_{-1.223}$ & $3.09^{+2.16}_{-2.25}$ & $5.71^{+5.71}_{-5.72}$ \\
WASP-167 & 765.9 & $2.68^{+0.91}_{-0.58}$ & $3.26^{+0.69}_{-0.62}$ & $5.83^{+5.74}_{-5.76}$ \\
\hline
\hline
\end{tabular}
\end{table}

\clearpage

\bibliography{PASPsample701}{}
\bibliographystyle{aasjournalv7}



\end{document}